\documentclass{aa}
\usepackage{txfonts}
\usepackage{natbib}
\usepackage{graphicx}
\bibpunct{(}{)}{;}{a}{}{,} 
\usepackage{longtable}
\usepackage[normal]{caption}

\begin{document}

   \title{Short-term variability of a sample of 29 trans-Neptunian objects and Centaurs}

\author{ A. Thirouin \inst{1}
   \and J.L. Ortiz \inst{1}
   \and R. Duffard \inst{1}
   \and P. Santos-Sanz \inst{1}
   \and F.J. Aceituno \inst{1}
   \and N. Morales \inst{1}
 }

\offprints{A.Thirouin \email{thirouin@iaa.es}}

\institute{$^1$ Instituto de Astrof\'{\i}sica  de Andaluc\'{\i}a,
CSIC, Apt 3004, 18080  Granada,  Spain }


\date{received/accepted}

\abstract{}{We attempt to increase the number of Trans-Neptunian objects (TNOs) whose
short-term variability has been studied and compile a high quality database with
the least possible biases, which may be used to perform statistical analyses.}
{We performed broadband CCD photometric observations using several
telescopes.} {We present results of 6 years of observations, reduced and
analyzed with the same tools in a systematic way. We report completely new
data for 15 objects (1998SG$_{35}$, 2002GB$_{10}$, 2003EL$_{61}$, 2003FY$_{128}$, 2003MW$_{12}$,
2003OP$_{32}$, 2003WL$_{7}$, 2004SB$_{60}$, 2004UX$_{10}$, 2005CB$_{79}$, 2005RM$_{43}$, 2005RN$_{43}$,
2005RR$_{43}$, 2005UJ$_{438}$, 2007UL$_{126}$ (or 2002KY$_{14}$)), for 5 objects we present a
new analysis of previously published results plus additional data
(2000WR$_{106}$, 2002CR$_{46}$, 2002TX$_{300}$, 2002VE$_{95}$, 2005FY$_{9}$) and for 9 objects we
present a new analysis of data already published (1996TL$_{66}$, 1999TZ$_{1}$,
2001YH$_{140}$, 2002AW$_{197}$, 2002LM$_{60}$, 2003AZ$_{84}$, 2003CO$_{1}$, 2003VS$_{2}$, 2004DW).
Lightcurves, possible rotation periods and photometric amplitudes are
reported for all of them. The photometric variability is smaller than
previously thought: the mean amplitude of our sample is 0.1mag and only
around 15\% of our sample has a larger variability than 0.15mag. The smaller
variability than previously thought seems to be a bias of previous
observations. We find a very weak trend of faster spinning objects towards
smaller sizes, which appears to be consistent with the fact that the smaller
objects are more collisionally evolved, but could also be a specific feature
of the Centaurs, the smallest objects in our sample. We also find that the
smaller the objects, the larger their amplitude, which is also consistent
with the idea that small objects are more collisionally evolved and thus
more deformed. Average rotation rates from our work are 7.5~h for the whole sample, 7.6~h for the TNOs alone and
7.3~h for the Centaurs. All of them appear to be somewhat faster than what
one can derive from a compilation of the scientific literature and our own
results. Maxwellian fits to the rotation rate distribution give mean values
of 7.5~h (for the whole sample) and 7.3~h (for the TNOs
only). Assuming hydrostatic equilibrium we can determine densities from our
sample under the additional assumption that the lightcurves are dominated
by shape effects, which is likely not realistic. The resulting average density is 0.92~g/cm$^3$
which is not far from the density constraint that one can derive from the apparent
spin barrier that we observe.} {}

\keywords{Solar System: Kuiper Belt, Techniques: photometric }

\maketitle

%

\section{Introduction}

The rotational properties of the asteroids
provide plenty of information about important physical properties,  such
as density, internal structure, cohesion and shape (e.g., 
\cite{Pravec-Harris2000,Holsapple2001, Holsapple2004}). As for asteroids, the rotational properties of the Kuiper
Belt Objects (KBOs) provide a wealth of knowledge about the basic physical
properties of these icy bodies (e.g. \cite{Sheppard-Jewitt2002,
Ortiz2006,Trilling-Bernstein2006,Sheppard2008}). In addition,
rotational properties provide valuable clues about the
primordial distribution of angular momentum, as well as the degree
of collisional evolution of the different dynamical groups in the
Kuiper Belt. Rotational properties can also provide empirical
tests of predictions based on models of the collisional evolution
of the Kuiper Belt \citep{Davies1997, Benavidez2009}.

Studies of short-term photometric variability of Kuiper Belt
Objects allow us to retrieve rotation periods from the
photometric periodicities  and these studies also provide
constraints on shape (or surface heterogeneity) by
means of the amplitude of the lightcurves. Therefore,
observational programs on time series CCD photometry are a good
tool for studing the Kuiper Belt. Unfortunately, most KBOs are faint
and CCD photometry programs are time expensive and require medium
to large telescopes. For this reason and in contrast to the case for asteroids,  the sample of KBOs for which short-term variability
has been studied is not large and more importantly, the sample of objects with known rotation periods
is severely biased toward large photometric amplitudes, the reason
being that the large amplitude objects produce lightcurves
that are much easier to observe and from which a rotation period
can be unequivocally obtained. Very long rotation periods
are also difficult to determine and scientists rarely publish null
results or failed attempts to derive lightcurves, which causes a
bias in the literature. In other words, the scientific literature in the Kuiper Belt
field consists mostly of high amplitude lightcurves, which are not
truly representative of the rotation properties of the whole
Kuiper Belt population.

Other biases are present in the literature, such as an overabundance
of large objects because larger bodies are usually
brighter and easier to observe. One can try to avoid this bias by studying
Centaur objects, which are not TNOs because they are not farther away than
Neptune but are widely accepted to originate in the Kuiper Belt; they
are thus KBOs that recently came to the inner solar system vicinity, with 
``recently" meaning time frames of several mega-years, the typical lifetime
of Centaurs \citep{Tiscareno2003}. The currently known Centaurs are
smaller than TNOs with the same brightness. Therefore, targeting Centaurs
provides information about small size objects of the Kuiper Belt that would
otherwise be too faint for telescopic studies.

Based on the aforementionned ideas we initiated a programme to photometrically monitor as many KBOs as
possible, trying to build a reasonably good sample in terms of
number of objects observed and also in terms of the amplitude
biases. Thus, we published even dubious rotation
periods of low-amplitude lightcurve objects rather than omitting
them. In their review of asteroid rotation rates, \cite{Binzel1989}
emphasized that excluding poor reliability objects results providing more weight to asteroids with large amplitudes and short
periods, introducing a significant bias.

We present new results from our survey and a
reanalysis of several bodies for which we had already published
results. We reanalyzed the data for these bodies either because we had acquired more
data or because we had developed superior analysis tools that resulted in what
we consider an improvement. All the objects presented here were
analyzed with the same tools and software and therefore represent
a homogeneous data set. The work presented here summarizes a
considerable effort in which more than 5000 images were
reduced and analyzed.

This paper is divided into 6 sections. Section 2 
describes the observations and the data sets. Section 3 describes
our software reduction tools and the methods used to derive e.g., 
periodicities, rotation periods and photometric range. Section 4
deals with the main results obtained for each object and Section 5
discusses the results altogether. Finally, our findings are
summarized in Section 6.

%

\section{Observations}

As already mentioned, our group at the Instituto de
Astrof\'{\i}sica de Andaluc\'{\i}a (IAA, CSIC) started a vast program on
lightcurves of the KBOs in 2001. Observations were carried out
from the 1.5~m telescope at Sierra Nevada Observatory (OSN -
Granada, Spain), from the 2.2~m CAHA telescope at Calar Alto
Observatory (Almeria, Spain), and from the 2.5~m Isaac Newton
Telescope (INT) at El Roque de Los Muchachos (La Palma, Spain).

The typical seeing during the observations at OSN ranged from 1.0"
to 2.0", with a median seeing around 1.4". The observations
reported here were carried out by means of a 2kx2k CCD with a
total field of view (FOV) of 7.8'x7.8'. We used a 2x2 binning mode,
which changes the image scale to 0.46"/pixel. This scale was
sufficient to ensure accurate point spread function sampling even in the
best seeing cases.

At Calar Alto, the median seeing of the observations was around
0.9". For our observations, we used the Calar Alto Faint Object
Spectrograph (CAFOS). This instrument is equipped with a 2048x2048
pixel CCD. Image scale is 0.53"/pixel, which is good enough to achieve
good sampling for most cases, although for the highest quality seeing some
images were slightly undersampled.

At INT, the median seeing was around 1$"$. Our observations were obtained
with the Wide Field Camera (WFC) instrument. This camera consists of 4
thinned EEV 2154x4200 CCDs. Image scale is 0.33"/pixel, which guarantees
good sampling even at the best seeing moments.

Exposure time had to be chosen by considering two main factors. On
the one hand, it had to be long enough to achieve a signal to noise
ratio (S/N) sufficient to study the observed object. On the other hand,
it had to be short enough to avoid elongated images of the target
(if the telescope was tracked at sidereal speed) or elongated
field stars (if the telescope was tracked at the KBO speed). We
always chose to track the telescope at sidereal speed. Our
observations concerned, basically, two kinds of objects: TNOs
and Centaurs. Drift rate of TNOs is typically low, $\sim$2"/h, so an exposure
time of around $\sim$300 to $\sim$600 seconds was used.The drift rate of Centaurs is higher than that of TNOs, typically being $\sim$10"/h. For Centaurs observations, we typically used an
exposure time of $\sim$200 seconds.

Observations were performed either unfiltered or using the Johnson Cousins
R filter to maximize the S/N. Among the
observations presented in this work, the majority were carried out
without filter, while the R filter was used for: 1999~TZ$_{1}$, Haumea
(2003~EL$_{61}$), Makemake (2005~FY$_{9}$), Orcus (2004~DW) and Varuna
(2000~WR$_{106}$) and for INT observations. Since the goal
of our studies is short term variability, we require relative photometry, not absolute photometry. Therefore, the use of unfiltered images
is not a concern for our work. Important geometric data of
the observed objects at the dates of observations analyzed here
are summarized in Table~1.

Our sample of objects was selected according to the their
brightnesses. Very faint objects cannot be observed with a 1.5~m or a 2~m
telescope with the needed S/N, thus we restricted our
target list to objects brighter than 21~mag in V as predicted magnitudes for
the dates of our observing runs according to the Minor Planet Center (MPC) ephemerides generator.

\section{Data reduction}

During each observing night, a series of bias and flat
fields were in general obtained to correct the images. We thus created a
median bias and a median flatfield for each day of observation. Care was
taken not to use bias or flat field frames that might be affected by observational
or acquisition problems. The median flatfields were assembled from twilight
dithered images and the results were inspected for possible residuals from
very bright saturated stars. The flatfield exposure times were always long
enough to ensure that no shutter effect was present so that a gradient or
an artifact of some sort could be present in the corrected images. Each
target image was bias subtracted and flatfielded using the median bias and
median flatfield of the observation day but if daily information about the bias
and flat field was not available, we used the median bias and median flat
field of a former or subsequent day. No cosmic ray removal algorithms were
used and we rejected the images in which a cosmic ray hit or a star was too
close to the object.

Relative photometry using as many as 25 field stars was carried
out by means of Daophot routines (\cite{Stetson1987}). The typical error bars of the
individual time integrations were $\sim$0.01~mag for the brightest
targets, and 0.06~mag for the faintest objects (in the poorest
observing conditions). Care was taken not to introduce spurious
results due to faint background stars or galaxies in the aperture.
If observations were adversely affected by cosmic-ray hits within
the flux aperture, we did not include them in our results.

We used a common reduction software for the photometry data
reduction of all the images, but since they came from three
different observatories, some parameters of the software were
specific for each data set. Those parameters were related to the
aperture size, which ranged from 6 to 24 pixels.

The choice of the aperture diameter is important. We had to choose
an aperture as small as possible to obtain the highest
signal to noise ratio but large enough to include most of the flux.
We typically used an aperture radius of the same order as the full
width at half maximum (FWHM) of the seeing. We carried out the
aperture photometry with several aperture radii values around the
FWHM. We also used an adaptable aperture radius, which was
different for each image and resulted from the fit of a Moffat
star profile to several stars in the field; this allowed us to
adapt to the varying seeing during the night. For all apertures
used, we chose the results that gave the lowest scatter in the
photometry of both the target and of stars with
similar brightness as our target. Several sets of reference stars
were used to obtain the relative photometry of all the targets,
and only the set that gave the lowest scatter was used. In many
cases, several stars had to be rejected from the analysis because
they exhibited some variability. The final photometry of our targets
was computed by taking the median of all the lightcurves obtained
with respect to each reference star. By applying this technique,
spurious results were eliminated and the dispersion of the
photometry improves.

During an observational campaign, we tried to keep the same
field and therefore the same reference stars. In some cases,
owing to the drift of the observed object, the field changed
completely or partially. If the field changed completely, we used
different reference stars for two or three subsets of nights in
the entire run. If the field changed partially, we tried to keep
the greatest number of reference stars in common during the whole
campaign. This number varied from 6 to 25. We generated TIFF
images within which all the reference stars were marked for each observation
night.

When we combined several observing runs, we had to normalize the
photometry data to its average because we did not have absolute
photometry that would allow us to link one run with the other;
in several instances, we did experiment with trying to
link several runs by using absolute photometry, and the errors
involved were generally much larger than what we can achieve by
normalizing the photometry to the mean or median value.
Furthermore, the small jumps in the photometry caused by the 
inevitable absolute photometry offsets cause spurious frequencies
in the periodogram analysis. This is especially true for very low
variability objects, which are numerous.  By normalizing the means of several runs, we assume that a similar 
number of data points are in the upper part and lower part of the curves. This may not be true if the runs are only two or three
nights long, but this is not usually the case. We emphasize that we
normalized the mean of each run not the mean of each night.

The final time series photometry of each target was inspected for
periodicities by means of the Lomb technique \citep{Lomb1976} as implemented
in \cite{Press1992}, but we also verified the results by using several other
time series analysis techniques (such as PDM), the \cite{Harris1989} method
and the CLEAN technique \citep{Foster1995}. As mentioned before, the
reference stars were also inspected for short-term variability. We can thus
be confident that no error has been introduced by the choice of reference
stars. To measure the amplitudes of the short-term variability, we performed
Fourier fits to the data to determine the peak to valley amplitudes
(full amplitudes).

As mentioned in the Section 2, we generally acquired our observations without a filter.
Using no filter may be a problem in some cases depending on sky conditions and CCD types.
Many CCDs have strong fringing effects caused by near-infrared
interference and  mainly related to e.g., their pixel size and thinning. But this was not the case for our observations.
By obtaining unfiltered images, we reached deeper magnitudes with sufficient signal to noise. We used an R filter when we anticipated that using no filter could be a problem. 

In some cases, we combined data obtained without a filter with data obtained with the R filter,
for example for 2005~RN$_{43}$ or 2005~RR$_{43}$.
We assumed that only the lightcurve amplitude is affected when one observes in a different filter, but the rotation period
would be the same. Besides, our unfiltered observations are close to R because the CCD sensitivity is usually reaches a maximum
at red wavelengths. Thus, we expect this effect to be small, but might alter the periodogram to some degreee. 

Even though absolute photometry was not the goal of the
observations, we computed approximate magnitudes for a few images
per object per observing run.
To obtain approximate R magnitudes, we used USNO-B1 stars in the field of view as photometric references. Since the USNO-B1 magnitudes
are not standard BVRI magnitudes and because we also did not use BVRI filters, we derived very approximate magnitudes, with a typical uncertainty of
0.4~mag.

The time series photometry of all the objects is provided as online Table 1 in the "Center of astronomical Data of Strasbourg" (CDS) and on the link: "www.iaa.es/$\sim$ortiz/thirouinetal2010/supportingonlinematerial.pdf" 
We have highlighted in bold face the times corresponding to the images that we used
to obtain an R magnitude calibration. The remaining R magnitudes in the table were obtained by using the relative magnitude information.
In online Table 1, we also present are geometric data such as geocentric and heliocentric distance and phase angle. The R magnitudes that the TNOs would have if they were at 1~AU from the Earth and the Sun (m$_\mathrm{R}$(1,1)) are also shown. No phase corrections were applied.

%

\section{Photometric results}

We present our results in terms of the following
classification of KBOs, which are based on dynamical criteria: (i)
\textit{\textbf{Classical group}} representing objects under the influence of Neptune
and away from the main mean motion resonances; (ii) \textit{\textbf{Resonant
objects}} which are in mean motion resonances with Neptune;
(iii) \textit{\textbf{Scattered Disk Objects (SDOs)}} which have a high
orbital eccentricity and have had a close encounter with  Neptune in the past
that sent them to their present position; and (iv) \textit{\textbf{Centaurs}} considered to be similar to TNOs but to reside between Jupiter's and Neptune's orbits, after having
experienced a close encounter with Neptune. We used the minor planet center
lists to classify all these bodies.

The lightcurves and Lomb periodograms for all objects are provided
as online material (online Fig 1 to online Fig 59) on the link: "www.iaa.es/$\sim$ortiz/thirouinetal2010/supportingonlinematerial.pdf" 
We only present an example of lightcurve in Fig 1.

\subsection{Classical objects}

\textit{\textbf{(120132) 2003~FY$_{128}$}} is a classical object observed on 09,
10, 11, 12 February 2005 and 09 March 2005 at the OSN telescope. The Lomb
periodogram (online Fig 1) for our data contains an important peak (confidence level $>$99\%)
at 8.54~h (2.81~cycles/day), which  is a single peaked periodicity. A
double peaked periodicity of 17.08~h might be more appropriate because the fit to
a Fourier series shows minima and maxima of different values, but neither PDM
nor the Harris method, which are less sensitive to the exact shape of the
lightcurve, proposed a periodicity 17.08~h (Fig 1 and online Fig 2). The amplitude is
0.15$\pm$0.01~mag. A second high peak in the periodogram of a lower
spectral power is located at 1.76~cycles/day, which appears to be an alias.

\cite{Sheppard2007} observed this object in the R band on 09, 10 March 2005 at
the Dupont 2.5~m telescope in Las Campanas in Chile. They presented a very
flat lightcurve based on 17 data points. They noted that 2003~FY$_{128}$ has no
significant short-term variability. A small amplitude $<$0.08~mag was
suggested. \cite{Dotto2008} observed this object on 18, 19 April 2007 at the 3.0~m New Technology
Telescope (NTT) at the European Southern Observatory at La Silla (Chile).
Their photometry appears to be inconsistent with some slight short-term
variability. They presented more than 13~h of observations, in R band but could not determine a rotational period. Their own evaluation suggests
a short-term variability longer than 7~h.

In conclusion, we present the first lightcurve of this object.
A period longer than 7~h proposed by \cite{Dotto2008} is consistent with
our results.

\textit{\textbf{(174567) 2003~MW$_{12}$}} is a classical object. This body was
observed on 28 May 2006, on 05, 06, 07, 10, 23, 24 June 2006 and on
12, 13, 14, 26, 27 April 2008 at the OSN telescope. We present two possible
lightcurves (online Fig 3), one based on a single peak
periodicity of 5.9~h (4.07~cycles/day) and another single peaked one of
7.87~h (3.04 cycles/day), which appears to be an alias. The Lomb periodogram (online Fig 4) suggests a periodicity of
5.9~h, but the Harris, PDM and CLEAN techniques infer a 7.87~h period. The
amplitude of the curve is 0.06$\pm$0.01~mag. From visual inspection, the best-fit lightcurve is obtained for a period of 5.9~h because the alternative fits exhibit more
scatter. To our knowledge, there is no published photometry of this object
to compare with. We present the first lightcurve of 2003~MW$_{12}$ and
propose a periodicity of 5.9~h or 7.87~h for this object.

\textit{\textbf{(120178) 2003~OP$_{32}$}} is a classical object observed on 05,
06, 07, 08, 10 March 2005, on 03, 04, 05 October 2005 and on 15, 16, 17
September 2007 at the OSN telescope. We propose a 0.13$\pm$0.01~mag
amplitude lightcurve with a short-term variability of 4.05~h
(5.93~cycles/day) (online Fig 5). A double
peaked periodicity of 8.1~h is neither supported nor excluded by the data. The
Lomb periodogram (online Fig 6) shows another high peak at 6.97~cycles/day and a second
peak smaller than the first one (but with a high confidence detection level) at
4.89~cycles/day. They both appear to be aliases of the main periodicity. All
techniques used confirm a periodic signature at 4.05~h.

We found one bibliographic reference for this target: \cite{Rabinowitz2008}
presented the results of R band observations carried out in 2006 July 17 - 2006
November 23 at the 1.3~m SMARTS telescope. Seventy-eight sparse sampled data points
formed a 4.845~h lightcurve with an amplitude of 0.26~mag.

Our results appear to be of much higher photometric precision than those of
\cite{Rabinowitz2008} as our lightcurve has a far smaller scatter and is based on
147 data points. The large 0.26~mag amplitude is rejected by our data, which
were optimized for period determination rather than for phase coefficient
determination (the primary goal of the sparse sampled observations by
\cite{Rabinowitz2008}).

\textit{\textbf{(120347) 2004~SB$_{60}$}} is a classical binary object.
The magnitude difference between
this object and its satellite is 2.3 (\cite{Noll2006}).
This object was observed on 05, 06, 07, 10 August 2005 and on 03, 04, 06, 07 August 2008 at the OSN
telescope. We present (online Fig 7) two lightcurves: a single
peaked periodicity of 6.09~h (3.94~cycles/day) or 8.1~h (2.96~cycles/day).
An amplitude of 0.03$\pm$0.01~mag is seen in both cases. Double peaked
lightcurves at 12.18~h or 16.02~h period do not appear to be more distinctive. PDM, CLEAN
and Lomb techniques (online Fig 8) favor a periodicity of 6.09~h, but the Harris method
detects a period of about 8.1~h.

To our knowledge, there is no bibliographic source on 2004~SB$_{60}$ rotational
variability. We present the first lightcurve of this body. As in the case of 
2003~MW$_{12}$, we cannot clearly favor a rotational period of either 6.09~h or
8.1~h. For very low amplitude objects, we know that very small night to night
offsets in the photometry can transfer a lot of power from the main peak to
a 24h-alias and viceversa. Therefore, even though 6.09~h is the preferred
result, 8.1~h is also quite possible and that could be true for other
aliases.

\textit{\textbf{2005~CB$_{79}$}} is a classical object. This object was observed
on 06, 07 January 2008 at the Calar Alto telescope, on 01, 04 May 2008 and
26 December 2008 at the OSN telescope. A single peaked periodicity of
6.76~h (3.55~cycles/day) is indicated by the Lomb periodogram (
online Fig 9). The lightcurve has an amplitude of 0.13~mag (online Fig 10). A double
peaked periodicity of 13.52~h is not preferred by any of the time series
analysis methods. Apparent aliases are at 2.55~cycles/day and at
4.55~cycles/day. All the techniques measure the same periods.

\textit{\textbf{(145452) 2005~RN$_{43}$}} is a classical body observed on 22
October 2006 at the INT and on 14, 16, 17, 19 September 2007 and 03, 04, 05, 07, 08
August 2008 at the OSN telescope. The Lomb periodogram (online Fig 11) exhibits a high peak
with a $>$ 99\% confidence level at 5.62~h (4.28~cycles/day) and a second
peak with a lower confidence level at 7.32~h (3.27~cycles/day). The
lightcurve has an amplitude of 0.04$\pm$0.01~mag (online
Fig 12). PDM,CLEAN and the Harris techniques confirm that the peak is located at
7.32~h. We cannot conclusively favor one period over the other for similar
reasons as for 2004~SB$_{60}$, but 7.32~h appears the most likely.

\textit{\textbf{(145453) 2005~RR$_{43}$}} is a classical object observed on 22,
23, 26 October 2006 at INT, on 15, 16, 17, 18 December 2006, on 11, 12, 13,
14, 15, 16 January 2007 at the Calar Alto telescope and on 14, 15, 17
September 2007 at the OSN telescope. The Lomb periodogram (online Fig 13) exhibits several
peaks. We note a significant peak located at 7.87~h (3.05~cycles/day) and
a second peak of  a lower significance level at 6.35~h (3.78~cycles/day).
PDM identifies the same peaks at the same values and a third peak at 4.1~cycles/day. We present a lightcurve with a
single peak periodicity of 7.87~h
(3.05~cycles/day) (online Fig 14). A double peaked lightcurve at
15.74~h does not appear more likely than the 7.87~h period. The amplitude of
the periodic signal is 0.06$\pm$0.01~mag.

\textit{\textbf{(50000) Quaoar (formerly 2002~LM$_{60}$)}} is a binary classical
object. The magnitude difference between Quaoar and its satellite is 5.6$\pm$0.2~mag (\cite{Brown2007}). Quaoar was observed on 21, 22, 23 May 2003, on 17, 18, 19, 20, 21, 22 June 2003 at the OSN telescope.
The Lomb periodogram (online Fig 15) for this object shows one peak with a high spectral power corresponding to a periodicity at
8.84~h (2.72~cycles/day), but a double peaked lightcurve at 17.68~h provides a more probable fit  because there appear to be two maxima of different height.
Both, the PDM and CLEAN techniques  confirmed the single peaked period with a high
spectral power but the Harris method suggests the double peaked
periodicity. The lightcurves are shown in the online Fig 16 both
of which have an amplitude of 0.15$\pm$0.04~mag.

Our data were used in \cite{Ortiz2003b}, who inferred a
17.67883~h double peaked periodicity with a confidence level above
99.9~\% and an amplitude of 0.133~mag.
This object was later observed by \cite{Rabinowitz2007}. They
presented on 166 R band observations carried out using the 1.3~m
telescope of the Small and Moderate Aperture Research Telescope
System (SMARTS). \cite{Rabinowitz2007} suggested a 8.84~h single
peaked rotational lightcurve with an amplitude of 0.18~mag, which
is consistent with our results.

\cite{Lin2007} presented R band results based on the analysis of 57 images acquired in
03,04 June 2003 and using the Lulin one-meter Telescope (LOT).
They proposed a 9.42~h single peaked rotational lightcurve with a
confidence level higher than 99.9~\% and a $\sim$0.3~mag amplitude.
Such a high amplitude is inconsistent with both our results and
those of \cite{Rabinowitz2007}, which were obtained using larger telescopes
and data sets.

In conclusion, we propose a 17.68~h double peaked periodicity or a 8.84~h
single peaked periodicity.

\textit{\textbf{(20000) Varuna (formerly 2000~WR$_{106}$})} is a classical
object. Varuna was observed on 05, 07, 31 January 2005 and on 01, 09, 10
February 2005 at the OSN telescope. The Lomb periodogram (online Fig 17) for this object shows a very
clear peak with a high spectral power corresponding to a periodicity of 3.1709~h (7.57~cycles/day),
but a double peaked lightcurve for a period twice that value has smaller
scatter. PDM confirms the 6.3418~h period. The lightcurve is depicted in the
online Fig 18 and shows an amplitude of 0.43$\pm$0.01~mag.

\cite{Ortiz2003a} presented results of observations at the OSN 1.5m telescope on 08, 09 February 2002. In that work, a lightcurve with an amplitude of 0.41~mag and a 6.3436~h double peaked rotational (3.1718~h single peak) was proposed. Varuna was also observed by \cite{Farnham2001}, who presented results of observations carried out on January, March and September 2001. They proposed a 6.34~h double peaked rotational lightcurve
with an amplitude of 0.50mag. \cite{Sheppard-Jewitt2002} used R-observations made on 17-18-19-20
February and April 2001 at the 2.2~m University of Hawaii
telescope. They suggested a 6.34~h double peaked rotational
lightcurve with a 0.42~mag amplitude. \cite{Belskaya2006} presented R-observations carried out on November,
December 2004 and on January, February 2005. Observations were made at 1.5~m
OSN telescope and at the bulgarian 2~m Ritchey-Chretien-Coude telescope.
They obtained a 6.34358~h double peaked rotational lightcurve with an
amplitude of 0.42~mag for phase angles larger than 0.8$^\circ$ and
0.47~mag near opposition.
In \cite{Belskaya2006}, it was shown that the lightcurve amplitude of Varuna
increases at small phase angles.

\cite{Rabinowitz2007} observed Varuna on 30 December 2004 and 17
April 2005. They presented photometric results based on 78 images.
They suggested a 6.344~h double peaked rotational lightcurve with
an amplitude of 0.49~mag.

Even though most of the works are consistent with a 0.42~mag amplitude lightcurve, the works by \cite{Farnham2001} and
\cite{Rabinowitz2007} measured a larger amplitude. However, \cite{Farnham2001} does not quote the uncertainty in his derivation.
On the other hand, \cite{Rabinowitz2007} data were not optimized for short-term variability studies.
The differences might therefore arise because of the different quality of the data, but since part of the small discrepancies in the amplitude of the lightcurve
reported by the various authors could be due to the phase angle variability reported in
\cite{Belskaya2006}, we recall that the observations of \cite{Ortiz2003a} were taken at phase
angles between $\alpha$~=~0.81~$^\circ$ and 0.83~$^\circ$, those by \cite{Farnham2001} covering the range $\alpha$~=~0.57~$^\circ$~-~0.63~$^\circ$, those by \cite{Sheppard-Jewitt2002} being taken at around 1~$^\circ$ in February 2001
and around 1.2~$^\circ$ in April 2001, those by \cite{Belskaya2006} being taken between $\alpha$~=~0.056~$^\circ$~and 0.92~$^\circ$ and those by \cite{Rabinowitz2007} covering the range $\alpha$~=~0.06~$^\circ$~-~1.3~$^\circ$. In conclusion, there is little question that Varuna has a double
peaked lightcurve and a 6.34~h rotational period with
0.42$\pm$0.01~mag amplitude, outside the opposition surge.

\textit{\textbf{(55565) 2002~AW$_{197}$}} is a classical object observed on 01-02
February 2003 and on 19, 21, 22, 23, 24, 25 January 2004 at the OSN
telescope.

The Lomb periodogram (online Fig 19) of this object contains several peaks with significant
confidence levels. The highest peak is located at 8.78~h (2.74~cycles/day).
The PDM, CLEAN and Harris techniques favored a rotational period of 8.78~h.
The lightcurve of this object (online Fig 20) has a very small
amplitude 0.04$\pm$0.01~mag. Thus it is difficult to judge whether a double
peaked lightcurve would be better.

\cite{Ortiz2006} published a possible 2002~AW$_{197}$ lightcurve. They used
observations of 29 November 2002, 01, 03, 05, 07, 08 December 2002, 01, 02,
03 February 2003 and 21, 22, 23, 25, 26 January 2004. \cite{Ortiz2006}
presented a Lomb periodogram with several high peaks (confidence level
$>$99.9$\%$). The highest peak was located at 8.86~h and was accompanied by
two aliases, with lower spectral power than the first peak, at 13.94~h and
at 6.94~h. Another peak with high confidence was identified at 15.82~h. Ortiz
et al. inferred, a 8.86~h period for a single peaked rotational lightcurve with a
reliability code of 2 (reliability code according to the definition
given in \cite{Lagerkvist1989}).

\cite{Sheppard2007} observed this body in the R band on 23, 24 December 2003
at the University of Hawaii 2.2~m telescope. They presented a very flat
lightcurve based on 27 data points. They noted that 2002~AW$_{197}$ has no
significant short-term variability. A small amplitude $<$0.03~mag was
suggested.

In conclusion, 2002~AW$_{197}$ has a nearly flat lightcurve and we propose a 8.78~h rotation period, although one should keep in mind that for the very
low amplitude objects many of the apparent 24~h aliases are also probable.

\textit{\textbf{(55636) 2002~TX$_{300}$}} is a classical object that was
observed on 07, 08, 09 August 2003 at the OSN telescope.

The Lomb periodogram (online Fig 21) identifies a single peak periodicity of 4.08~h
(5.88~cycles/day) but PDM and the Harris method determine the true periodicity to be twice that value (8.16~h), with a 0.04$\pm$0.01~mag amplitude
for the 4.08~h period and 0.08$\pm$0.01~mag for the 8.16~h period.
Online Fig 22 shows the lightcurve for 8.16~h.

\cite{Ortiz2004} presented observations performed on 29, 30 October 2002, on 06, 07,
28, 29, 30 November 2002 and on 02, 03, 04, 06 December 2002. Coordinated
observational campaigns were arranged at three different telescopes: the
1.5~m OSN telescope, the 3.6~m Canada-France-Hawaii Telescope (CFHT) and the
2.5~m Nordic Optical Telescope (NOT) at La Palma for the photometric study.
Results of this coordinated campaign showed a 7.89~h peaked rotational
lightcurve with an amplitude of 0.09~$\pm$0.08~mag.

Here, we used additional images from two nights that were not available in \cite{Ortiz2004}.
In Ortiz et al. the two larger aperture telescopes provided  a few data
that could be used to check whether the 7.89~h was consistent with those observations or not.
On the other hand, data from the larger aperture telescopes were scarce and of poorer quality than the data from the 1.5~m OSN telescope.
Therefore, we consider 8.16~h to be more probable than 7.89~h. 

\cite{Sheppard-Jewitt2003} presented a 8.12~h or a 12.1~h single
peaked rotational R band lightcurve with an amplitude of
0.08~$\pm$0.02~mag.

Thus, our results and those of Sheppard's are consistent and not far from
the \cite{Ortiz2004} results. This implies that a rotation period of
8.14$\pm$0.02~h with a photometric range of 0.08$\pm$0.02~mag is secure.

\textit{\textbf{(55636) Makemake (formerly 2005~FY$_{9}$)}} is a classical
object. This object was observed on 07, 08, 10, 12 April 2005, on 27-29 May
2006, on 05, 07, 10 June 2006, on 14-18 December 2006 and on 09-12 March 2007
at the OSN telescope and on 11-16 January 2007 at Calar Alto telescope. This
object was extensively observed because it had an extremely low (and
challenging) variability. The Lomb periodogram (online Fig 23)
shows two broad maxima containing many sharp spikes. The two maxima
correspond to $\sim$7.7~h ($\sim$3.1~cycles/day) and its 24-h alias at
$\sim$11.5~h (2.1~cycles/day). The period of 7.7~h is slightly more likely in terms of
spectral power, but the exact spike at which the lightcurve is more accurately fit is not
straightforward to find. We chose 7.64~h and the corresponding lightcurve is
shown in the online Fig 24. With fewer data, \cite{Ortiz2007} favored what we understand to be the 24-h alias. As mentioned before, much spectral power is easily transferred from one peak to its 24-h alias in low
variability objects, for which 2005~FY$_{9}$ is the most extreme case that we
deal with in this paper. Very slight night to night calibration errors are
the reason for the transfer of spectral power. Thus, it is difficult to
decide which is the correct period. However, since we have a large number of
nights, random night-to-night offsets should cancel on average. Thus, we
believe that 7.7~h is more likely than the value that was derived with fewer
nights in \cite{Ortiz2007}. The amplitude of the lightcurve obtained by
means of a sinusoidal fit is 0.015$\pm$0.005~mag.

\subsection{Resonant objects}

\textit{\textbf{(126154) 2001~YH$_{140}$}} was observed on 15, 16, 17, 18, 19
December 2004. This object is in resonance 3:5 with Neptune. The Lomb
periodogram (online Fig 25) and PDM show three peaks of high spectral power located at
13.2~h (1.82~cycles/day), 8.40~h (2.86~cycles/day) and 6.19~h
(3.88~cycles/day). The lightcurve for a rotation period of 13.2~h is shown
in the online Fig 26. The amplitude of the lightcurve is
0.13$\pm$0.05~mag.

Using data from 15-20 December 2004, \cite{Ortiz2006} favored a
periodicity of 8.45~h but, presented two aliases located at
6.22~h and at 12.99~h. \cite{Sheppard2007}, from R band
observations, taken on 19, 21, 23, 24 December 2003 at the University
of Hawaii 2.2~m telescope, suggested a double peak periodicity of
13.25~h with an amplitude of 0.21~$\pm$0.04~mag.

In summary, there is agreement about the rotation period of this body, bearing in mind that 13.20~h is very close to the 12.99~h possible
alias reported in \cite{Ortiz2006}. However, the amplitude that we report is
somewhat different, although consistent within their error bars.

We propose a lightcurve with a single peak periodicity of 13.2~h
and an amplitude of 0.13$\pm$0.04~mag, which appear to be
consistent with the data by \cite{Sheppard2007}.

\textit{\textbf{(90482) Orcus (formerly 2004~DW)}} is a binary object
in resonance 2:3 with Neptune. The magnitude difference between
this object and its satellite is 2.54$\pm$0.01~mag (\cite{Brown2009}).
Our observations were carried out on 08, 09, 11, and 23
March 2004 and on 22, 23, 25, 26, 27 April 2004 at the OSN telescope. The Lomb
periodogram (online Fig 27) and PDM technique show one peak with a high significance level
($>$99\%) located at 10.47~h (2.29~cycles/day). We propose a lightcurve with
this period in the online Fig 28 with an amplitude of
0.04$\pm$0.01~mag. All techniques PDM, CLEAN and Harris suggest the same
period of 10.47~h.

\cite{Ortiz2006} presented observational results on 08, 09, 10, 11
March 2004 and on 22, 23, 25, 26, 27 April 2004. The work published by
Ortiz and the present work used almost the same data but the
present work considers data for two more nights. The Lomb periodogram showed
three peaks with a high confidence levels ($>$99.9\%) located at
7.09~h, 10.08~h, and 17.43~h. The 7.09~h and 17.43~h periods were
probably aliases of the 10.08~h periodicity that appeared most likely. Applying different techniques such as PDM, CLEAN and chi- square
minimization, etc., the value 10.08~h was favored with a
reliability code of 2 (reliability code according the definition
given in \citep{Lagerkvist1989}).

\cite{Sheppard2007} observed Orcus on 14, 15, 16 February 2005 and
on 09, 10 March 2005 with an R-filter. His results were based on 43
images carried out at the Dupont 2.5~m telescope at Las Campanas
in Chile. He did not notice a significant short-term variability.
Lightcurves of each observational night suggested a flat
lightcurve with an amplitude $<$0.03~mag.

\cite{Rabinowitz2007} presented photometric results of 143 images
carried out on 21 February to 6 July 2004 using the 1.3~m SMARTS
telescope. They suggested a 13.19~h single peaked rotational
lightcurve with an amplitude of 0.18~mag. This is obviously
inconsistent with both \cite{Sheppard2007} and our work, in terms of both period and amplitude.

In conclusion, we present a lightcurve based on 327 images for which we
found a periodicity of 10.47~h, similar to \cite{Ortiz2006}.
Because of the small amount of data from \cite{Sheppard2007}, no
period was reported. Since the object has a very low amplitude
some of the 24~h aliases (at 7.3~h, 5.6~h and 18.54~h) could be the
true rotation too.

\textit{\textbf{(555638) 2002~VE$_{95}$}} was observed on 19 January 2004 and on
15, 16, 17, 18 and 19 December 2004 at the OSN telescope. This object is in
resonance 2:3 with Neptune. The Lomb periodogram (online Fig 29) for this object and PDM do not allow
us to identify a clear rotational periodicity. We note a high peak at
9.97~h (2.41~cycles/day) and three aliases at 17.32~h (1.39~cycles/day),
6.18~h (3.88~cycles/day) and 4.90~h (4.90~cycles/day). We obtain a lightcurve
(online Fig 30) with a 9.97~h periodicity and an amplitude of
0.05$\pm$0.01~mag.

\cite{Ortiz2006} presented observations carried out on 29 November
2001, 03, 04, 08 December 2002 and on 15, 16, 17, 18, 19, 20 December
2004. The Lomb periodogram showed three high peaks with a
confidence level $>$99.9\% located at 6.76~h, 7.36~h and 9.47~h.
Studying the data with other techniques, such as PDM, a 6.88~h high
peak was detected. The CLEAN technique confirmed the peak detected
by PDM. \cite{Ortiz2006} did not favor or discard any
periodicity. In all cases, the amplitude of the lightcurve was
below 0.08~mag.

This object was also observed by \cite{Sheppard-Jewitt2003}. They
presented results of observations made on the University of Hawaii
2.2~m telescope during one night. They could not determine a
periodicity. An amplitude variation of $<$0.06~mag was detected
during their observations.

In conclusion, even though the periodicity that we detect is above the
99\% confidence level, the periodicityis probably caused by small instrumental errors that are difficult to evaluate.
One should keep in mind that 2002~VE$_{95}$ has
very low amplitude variations and the overall scatter of our data
is greater than for other low variability objects that
we have studied in more detail. Since \cite{Ortiz2006} did not identify any periodicity and \cite{Sheppard-Jewitt2003} could not determine a
period, our derivation is only tentative and more
observation data with smaller scatter will be necessary to derive a rotation period completely reliable.

\textit{\textbf(208996) {2003~AZ$_{84}$}} is a binary object in resonance 2:3 with Neptune.
The magnitude difference between this object and its satellite is 5.0$\pm$0.3~mag (\cite{Brown2007}).
This object was observed on 22, 23, 24, 25 January 2004 and on 14, 15, 16, 17, 18, 19 December 2004 at the OSN telescope. The Lomb periodogram (online Fig 31) has one peak located at 3.53~cycles/day (6.79~h). In the online Fig 32, we present a 6.79~h
single peaked lightcurve with an amplitude of 0.07$\pm$0.01~mag. This period
is also derived from the PDM analysis and both the CLEAN and Harris
methods.
\cite{Ortiz2006} presented data acquired on 20, 22, 25, 26 January 2004 and
15, 16, 17, 18, 19, 20 December 2004, the Lomb periodogram of which exhibited several
peaks of a very high confidence level. The highest peak was located at
5.28~h and presented two aliases of similar spectral power at
4.32~h and at 6.76~h.

\cite{Sheppard-Jewitt2003} studied three observation nights:
23, 24, 25 February 2003 using the University of Hawaii 2.2~m
telescope. They presented a 6.72~h single peaked rotational
lightcurve with an amplitude of 0.14~mag.

Thus, we conclude that a rotation period is 6.75$\pm$0.04~h (a double peaked lightcurve at 13.5~h is apparently no
more likely than 6.75~h).

\textit{\textbf{(84922) 2003~VS$_{2}$}} was observed on 22, 26, 28 December 2003,
on 04, 19, 20, 21, 22 January 2004 at the OSN telescope. This object is
in resonance 2:3 with Neptune. The Lomb periodogram (online Fig 33) and PDM technique show a clear main
peak with a high confidence level ($>$99\%). We propose a short-term
variability of 3.71~h (6.47~cycles/day), but the double peaked version
(7.42~h) is more appropriate and corresponds to a clearer lightcurve with the
two maxima and two minima of different values (online Fig 34).
An amplitude of 0.21$\pm$0.01~mag is derived from our data.

\cite{Ortiz2006} presented data reduction of observations carried
out on 22, 23, 26, 28, 29 December 2003 and 21, 22 January 2004. The
Lomb periodogram showed two peaks with a high significance level
at 3.71~h and 4.39~h. The second value was an alias of the first
value. The amplitude was estimated as 0.23~mag.

Our results are also consistent with \cite{Sheppard2007}, who
presented a study of observations made with the University of Hawaii
2.2~m telescope. Observations were carried out on 19-21-23-24
December 2003 with an R-filter. They proposed a lightcurve with a
double peak periodicity of 7.41~h with an amplitude of 0.21~mag.

In conclusion, there is clear evidence that a 7.42~h
periodicity of 2003~VS$_{2}$ is its true rotation period. The
lightcurve presents an amplitude of $~$0.21~$\pm$0.01mag.

\textit{\textbf{(144897) 2004~UX$_{10}$}} was observed on 14, 17 September 2007
and on 30 November 2007 at the OSN telescope. This object is in
resonance 2:3 with Neptune. The Lomb periodogram (online Fig 35) shows three peaks with a high
confidence level ($>$99\%) at 4.23~cycles/day (5.68~h), at 3.88~cycles/day,
and at 4.58~cycles/day. The highest peak is 4.23~cycles/day and the other
values resemble aliases. However, all of them have very similar spectral
power. PDM determines a periodicity of 5.30~h. This value is consistent with
the Lomb one. We propose (online Fig 36) a lightcurve with a
short-term variability of 5.68~h (4.23~cycles/day). A double peaked
periodicity of 11.36~h does not give produce a closer fitting lightcurve and
the amplitude of 0.08$\pm$0.01~mag is indicated by the fits.

To our knowledge, we present the first photometric study of 2004~UX$_{10}$ so
there is no literature source with to compare our results which. From our results,
a periodicity of 5.68~h is possible, but because of the presence of
aliases and the low amount of data relative to those for other objects, it is safer to
conclude that the rotation period is between 5 and 7 hours. More observations would be needed to derive a more precise period for this
object.

\textit{\textbf{(136108) Haumea (formerly 2003~EL$_{61}$)}} is considered
as a resonant object by \cite{Ragozzine2007} at the 7:12 resonance with Neptune, although it was listed as belonging to the classical type by the Minor Planet Center
for quite some time. The magnitude differences between Haumea and its satellites Hi'iaka and Namaka, are 2.98$\pm$0.03~mag and 4.6 mag (respectively) (\cite{Lacerda-Jewitt2008}).
We present results of R band
observations on 12, 13, 14, 15, 16, 17 January 2007 at the 2.2~m Calar Alto
telescope. We obtained a double peaked lightcurve with a periodicity of
3.9153~h (single peak periodicity of 1.9577~h (12.24~cycles/day) indicated
by the Lomb periodogram (online Fig 37)), but both the PDM and Harris methods found the double peaked periodicity. The amplitude of our lightcurve is
0.28$\pm$0.02~mag (online Fig 38).

\cite{Rabinowitz2006} presented R-observations acquired on 25
January-26 July 2005 using the 1.3~m SMARTS telescope. They used
additional observations on 10-12 April 2005 at the 5.1~m Hale
telescope at Palomar Observatory and on 4 May 2005 at the 0.8~m
telescope at Tenagra Observatory. They derived a 3.9154~h double peaked
rotational lightcurve with an amplitude of 0.28~mag from their data.

\cite{Lacerda-Jewitt2008} used observations carried out on
11, 13, 15 June 2007 and on 07, 08, 22, 24 July 2007 at the 2.2~m
University of Hawaii telescope. Observations wereperformed using the R, B,
and J filter. They obtained a 3.9155~h double peaked rotational
lightcurve with an amplitude of 0.29~mag.

Thus, there is little doubt that 3.9154$\pm$0.0001~h is the correct
rotation period, as there is wide agreement on this from three
independent data sets. The same is true for the amplitude of the
variability.

\subsection{Scattered Disk Objects (SDOs)}

\textit{\textbf{(145451) 2005~RM$_{43}$}} was observed on 13, 14 October 2006,
15, 17, 18 December 2006 at the OSN telescope and on 11, 12, 13, 14, 15 January
2007 at the Calar Alto telescope. The Lomb periodogram (online Fig 39) exhibits a very high
peak at 3.58~cycles/day (6.71~h) and two aliases of this peak at 2.58 and
3.80~cycles/day. The lightcurve (online Fig 40) appears to have an
amplitude of 0.04$\pm$0.01~mag. PDM confirms
the first peak and shows these two other peaks with a lower confidence
level. We are therefore able to confirm a periodic single peaked lightcurve of
6.71~h (3.58~cycles/day) and a double peaked periodicity of 13.42~h. All other period finding techniques imply a periodicity of around $~$6.7~h.
There is no bibliographic reference with which to compare our
results.

\textit{\textbf{(42355) Typhon (formerly 2002~CR$_{46}$)}} is a binary object.
The magnitude difference between this object and its satellite is 1.30$\pm$0.06~mag (\cite{Grundy2008}).
Typhon was observed on 28 January 2003 and on 02, 04, 06, 09 March 2003 at the
OSN telescope. The Lomb periodogram (online Fig 41) shows several peaks, but one of them has a much higher spectral power. Thus, we present a lightcurve
corresponding to this periodicity in online Fig 42 that has a 9.67~h
(2.48~cycles/day) single peak period, a very small amplitude
0.07$\pm$0.01~mag. 24h-aliases are also
present. All techniquesinfer consider results.

\cite{Ortiz2003a} presented observations carried out on 08-10
March 2002. The Lomb periodogram showed two peaks, with a
confidence level below 50\%, located at 3.66~h and 4.35~h. Both
values were aliases. Because of a low significance level, no period
was favored. The amplitude was reported to be $<$0.15~mag.

\cite{Sheppard-Jewitt2003} observed Typhon during 4 nights using the
University of Hawaii 2.2~m telescope, but their study could not estimate a
periodicity. They presented a flat lightcurve with an amplitude $<$0.05~mag, which is consistent with our findings.

In conclusion, Typhon presents a nearly flat lightcurve, according to our
result and published articles. The period proposed in the present
work is tentative as we know that low variability objects are easily
affected by small night-to-night instrumental/observation changes that can
artificially accentuate the power of some spurious frequencies.

\textit{\textbf{(15874) 1996~TL$_{66}$}} was observed on 15, 16, 17, 18 December
2004 at the OSN telescope. The Lomb periodogram (online Fig 43) shows several peaks all of equally low
confidence. We cannot reliably determine a periodicity.
We are only able to identify the peak with the highest spectral power at 8.04~h
(2.99~cycles/day) and two aliases located at $~$12~h and $~$6~h. The
lightcurve presented in the online Fig 44 is a single peak
periodicity of 12~h with an amplitude of 0.07$\pm$0.02~mag. The CLEAN and the Harris analysis suggest a period of 5.1~h
and PDM propose 10.2~h.

Using data obtained on 14-19 December 2004, \cite{Ortiz2006} presented a 12.1~h single peaked rotational lightcurve with a
$<0.12$~mag amplitude. But according to the reliability code assigned
to this value by \cite{Ortiz2006} (code defined in
\cite{Lagerkvist1989}) this period is clearly uncertain.

\cite{Luu-Jewitt1998} used the 6.5~m Multiple Mirror Telescope
(MMT) on Mount Hopkins, Arizona to observed this object in the R band.
Observations were made during one night (over 6~h) on 15 October 1996. They
noted an amplitude $<$0.06~mag but they could not determine a periodicity.

This object was also observed by \cite{Romanishin-Tegler1999}
with the 2.3~m telescope on Kitt Peak, Arizona. Observations were
made between the 2nd and the 9th of October 1997 in V band.
Results were based on only 25 images. They could not identify a
short-term periodicity, but they noted an amplitude $<$0.06~mag.

In conclusion, the periodicity of this object is uncertain, there
being indications of 12~h and its diurnal aliases. There are also
indications of 10.2~h and even 5.1~h. With more observations, a more reliable
period might be secured.

\subsection{Centaurs}

\textit{\textbf{(52872) Okyrhoe (formerly 1998~SG$_{35}$)}} was observed on
05, 06, 07, 08, 10, 11, 12, 13, 14, 15 December 2006 at the OSN telescope. The Lomb
periodogram (online Fig 45) suggests a single peaked periodicity of 4.86~h (4.94~cycles/day)
or 6.08~h (3.95~cycles/day) and double peaked periodicities of 9.72~h or
12.16~h are also possible, but there is no clear evidence that they are any
better. Lightcurves present the 4.86h and the 6.08h periods (online
Fig 46). An amplitude of 0.07$\pm$0.01~mag is suggested from the fits.
The CLEAN method suggested a 6.08~h period, whereas the Harris method and PDM imply a 4.86~h. This is not surprising because both peaks in the periodogram have almost the same power.

\cite{Bauer2003} observed Okyrhoe at the University of Hawaii
2.2~m telescope. Observations were performed during 3 consecutive
nights from 22 to 24 September 1999, in R band. Their result was a
16.6~h double peaked rotational lightcurve with an amplitude of
0.2~mag. However, such a high amplitude is ruled out by our data
so we suggest that some kind of observational, instrumental, or
reduction problem affected their photometry.

In conclusion, 4.86~h and 6.08~h appear as possible values.

\textit{\textbf{(145486) 2005~UJ$_{438}$}} was observed on 11, 12, 13, 15, 16 January
2007 and 06, 07 January 2008 at the Calar Alto telescope and on 26 December
2008 at the OSN telescope. The Lomb periodogram (online Fig 47) shows several peaks with
high spectral power at 4.16~h (5.77~cycles/day). This is the highest peak
but there are important diurnal aliases. We propose a single peaked
periodicity of 4.16~h (5.77~cycles/day) with an amplitude of
0.13$\pm$0.01~mag (online Fig 48). CLEAN, Harris, and PDM all determine a
4.18~h rotational period.

There is no literature reference on photometric results for this
body that we are aware of. Thus we cannot compare our results with
others and our preliminary conclusion is that 4.18~h seems a
reasonable value with the caveats that the apparent 24-h aliases
can be the true periodicity.

\textit{\textbf{2002 KY~$_{14}$ = 2007~UL$_{126}$}} was observed on 01, 02, 03, 04, 05
August 2008 at the OSN telescope. The Lomb periodogram (online Fig 49) and PDM technique
show two peaks with a high spectral power located at 3.56~h
(6.74~cycles/day) and at 4.2~h (5.71~cycles/day). In both cases, the
lightcurve (online Figs 50 and 51) has an amplitude of
0.13~$\pm$0.01~mag. CLEAN, PDM, and Harris
suggest a 4.2~h rotational period. Thus we adopt this asour most likely estimate
of the rotation period. Double peaked versions of the two mentioned
periodicities do look slightly more probable but there is no
quantitative evidence to suppor this impression. In conclusion,
3.56~h and 4.2~h are possible rotation periods.

\textit{\textbf{(55567) Amycus (formerly 2002~GB$_{10}$)}} is a binary TNO observed
on 08, 09 March 2003 at the OSN telescope. The Lomb periodogram {online Fig 5}) shows two
peaks with a high spectral power at 2.46 and 1.48~cycles/day (9.76~h and
16.21~h respectively). The other period finding methods
suggest periods of between 9.7~h and 10.1~h. These peaks have a similar
spectral power, but 9.76~h is preferred. We propose a lightcurve
corresponding to this period. A single peak lightcurve of 9.76~h
(2.46~cycles/day) is presented in online Fig 53. The amplitude of
the variability is 0.16~$\pm$0.01~mag.

As far as we know, there is no bibliographic reference for the time
series analysis of Amycus that we can use to compare with and improve
our study. In conclusion, we can estimate a clear periodicity
around $\sim$10~h. More observations shouldl permit us
to determine a more precise rotational period.

\textit{\textbf{(120061) 2003~CO$_{1}$}} is a binary object observed on
19, 21, 22, 23, 24, 25 January 2004 and on 19, 23, 25, 26, 27 April 2004 at the OSN
telescope. The Lomb periodogram (online Fig 54) exhibits several peaks. We propose a lightcurve
with a single peak periodicity of 4.51~h (5.31~cycles/day) and an amplitude
of 0.07$\pm$0.01~mag (online Fig 55). Apparent aliases are at
4.3~cycles/day and 6.3~cycles/day. All methods confirm that the 4.51~h period is
the most likely choice.

The Lomb periodogram in \cite{Ortiz2006} proposed several
possibilities of a 3.53~h, 4.13~h, 4.99~h or 6.30~h rotational period.
\cite{Ortiz2006} favored a 4.99~h single peaked rotational
lightcurve. In all cases, the amplitude was 0.1~mag.

In conclusion, with the same observational data, our result and
\cite{Ortiz2006} result proposed somewhat different periods but once aliases
are taken into account, the closest agreement seems to be around the 5h range.

\textit{\textbf{(136204) 2003~WL$_{7}$}} was observed on 05, 06, 07, 08, 10, 11, 13, 14
December 2007 at the OSN telescope. The Lomb periodogram (online Fig 56), PDM, CLEAN, and the
Harris techniques suggest one main periodicity located at 8.24~h
(2.92~cycles/day). We propose a lightcurve based on that period
(online Fig 57). The amplitude is 0.05~$\pm$0.01~mag. A double
peaked periodicity of 16.48~h is not preferred by any criteria so we propose 8.24~h period as our most robust estimate.

There is no bibliographic reference of photometric results for
2003~WL$_{7}$ with which we can compare with. Based on 303 images, our result
appaears to be robust enough and 8.24~h seems a secure value, but the
low amplitude of the variations raises some concerns that one of
the diurnal aliases might be more appropriate.

\textit{\textbf{(12929) 1999~TZ$_{1}$}} was initially classified as a Centaur and
is still listed as such in the Minor Planet Center. \cite{Moullet2008} clearly demonstrated that 1999~TZ$_{1}$ is a jovian Trojan. We
should have chosen not to include this object in our study, but since the
object is still listed as a Centaur and because Trojans are also linked to
TNOs according to one of the dynamical models of the Kuiper Belt formation
and evolution \citep{Morbidelli2005}, its presence should not contaminate
our sample too much. In this work, we present data reduction of observations
carried out on 23-25 February 2007, on 09, 10, 11, 12 March 2007 at the OSN
telescope. The Lomb periodogram (online Fig 58) and PDM show one main peak with a confidence
level around 99\%. We propose a single peak periodicity of 5.211~h
(4.70~cycles/day) or a double peaked periodicity of 10.422~h with an
amplitude of 0.07$\pm$0.01~mag. In fact, the double peaked version appears a
slightly better option because the two minima of the lightcurve look
different by nearly 0.02~mag (online Fig 59). The Harris method
also favors this option.

Using the same data as used in our work, \cite{Moullet2008} presented a 10.438~h double peaked rotational lightcurve with an
amplitude$<$0.10~mag. This is entirely consistent with our own
analysis.

\cite{Dotto2008}, studied images carried out with the 3.5~m New
Telescope Tecnhology (NTT, Chile, La Silla). They observed this
object during one night (7~hours) and phased their data to
the period proposed in \cite{Moullet2008} to conclude that the
period in that paper was consistent with their observations.

\section{Discussion}

In Table 2, we summarize our results to allow us to perform an easier
interpretation of the data. In the online material, we show all the lightcurve plots
with the same vertical scale (relative magnitude) to make
comparisons between all the objects easier. One thing that is obvious
in the table and in the plots is the fact that most of the objects
present low amplitude variability. The average amplitude of the
variability in our sample is 0.1~mag. There are only 3 to 5 cases
(taking into account the error bars) in which the variability is greater than 
0.15~mag within our sample. This means that the percentage of
objects with high variability is between 10 and 20\%. This is
much smaller than previous estimates \citep{Jewitt-Sheppard2002,
Ortiz2003a, Ortiz2003b, Lacerda-Luu2006}, possibly because the
objects that were then reported were preferentially those for
which a clear periodicity could be derived, which usually requires
high amplitude lightcurves. This was already noted by \cite{Ortiz2003b}, who highlighted a possible overrepresentation of high amplitude objects. This possible bias was
also emphasized in the review by \cite{Sheppard2008}.

Low amplitude lightcurves are generally caused by albedo
heterogeneity on the surfaces of the bodies, although elongated
objects seen at certain geometries can also produce nearly flat
lightcurvesl. The physical reason for many
low amplitude rotators in the Kuiper Belt is investigated by \cite{Duffard2009}. The smallest amount of variability would be expected for MacLaurin spheroids with modest
to small surface heterogeneity. Hence, the high numbers of nearly
flat lightcurves might be indicative of many
MacLaurin shapes in the trans-neptunian region. A model to test
this and other ideas is presented in \cite{Duffard2009}.

According to our definition, we consider that
the limit to a high lightcurve amplitude is above 0.15~mag.
The high amplitude lightcurves of large objects which we can clearly
attribute to an aspherical shape can indicate the
typical magnitude of hemispheric albedo changes if we compare the
two maxima or two minima in the double peaked lightcurves. These
differences in the cases of 2003~VS$_{2}$ and Haumea are around
0.04~mag, whereas for Varuna the greatest difference is 0.1~mag.
Hence, this means that the hemispherically averaged albedo
typically has variations around 4 to 10\%. Thus we expect that the
variability induced by surface features is on the order of
0.1~mag.  For the asteroids, albedo variegations are usually
responsible for lightcurves amplitude between 0.10~mag and 0.20~mag at most
\citep{Magnusson-Lagerkvist1991}. We adopt here an
in-between value of 0.15~mag as the most reliable threshold
above which we can be nearly confident that the variations are caused by shape effects. This value has been used by several
investigators as the transition from low variability to
medium-large variability (e.g. \cite{Sheppard2008}).

A plot of rotation periods versus H parameter is shown in Fig. 2.
According to the plot, there is only a very slight indication
that objects with large H rotate faster. This trend is more evident
in \cite{Duffard2009}, where a larger sample is used.
Because H is a proxy for size, this implies that the smaller
objects rotate faster than the larger ones and that would be
consistent with the usual collisional scenario in which the small
objects are fragments and are more collisionally evolved than the
large objects \citep{Davies1997}. Since collisions tend to spin up
the bodies, the faster rotation rates for the smaller objects
seems to be consistent with this idea, but one should keep in mind
that the small objects studied here are all centaurs and they
might have suffered specific processes that could lead to spin up.

Two objects are rapid rotators: Haumea with a period of 3.92~h and
2003~OP$_{32}$ with a rotational periodicity of 4.05~h. Based on our
sample of data, there is an apparent spin barrier at between around 4~h
to 3.9~h. An object with a period shorter than this limit is
out of equilibrium.

The critical period $P_{c}$ is defined by equating the centrifugal
acceleration to the acceleration caused by gravity. From that
constraint, it follows that

\begin{equation}
P_{c} = \left(\frac{3\pi}{G\rho}_{c}\right) ^{\frac{1}{2}}
\end{equation}
where $G$ is the gravitational constant and $\rho_{c}$ the critical
density.

Since a rotational period of 3.90~h is the critical
rotational period, we can derive a lower limit to the density. The
result based on our sample suggests a lower limit to the density
of 0.71~$g/cm^{3}$. \cite{Davidsson1999, Davidsson2001} pointed
out that the critical period in Eq. 1 is not a reliable estimate for
true bodies and derived alternative expressions to Eq.1.
Using Davidsson's expression for a low tensile strength of
0.01~MPa and a radius of 100~km, we obtain a lower limit to the
density of 0.70~$g/cm^{3}$.

Another interesting subject is the plot of the amplitude of the
variability as a function of the H parameter for each object presented in Fig 3. One can immediately see the group of centaurs
at large Hs and their amplitudes are apparently systematically
above those from the regular TNOs (provided that the two very high
amplitude TNOs are not taken into account). The centaur population
lacks extremely low amplitude objects. In Fig. 3, two linear
fits are shown: the thick line is a fit based on all our data. The
dashed line is a linear fit to our data except for the objects
with a peak to peak amplitude $>$0.20~mag (Varuna, Haumea and
2003~VS$_{2}$). This fit clearly shows the trend of higher amplitude
toward larger H objects.

Since a TNO is a triaxial ellipsoid with axes
(a$>$b$>$c), one can determine that the lightcurve amplitude
$\Delta{m}$ varies according to the observational or viewing angle
$\xi$ (e.g., \cite{Binzel1989}) 

\begin{equation}
\Delta{m} = 2.5~log\left( \frac{a}{b}\right)  - 1.25~log\left(
\frac{a^{2}\cos^{2}\xi + c^{2}\sin^{2}\xi}{b^{2}\cos^{2}\xi +
c^{2}\sin^{2}\xi}\right)
\end{equation}
The maximum of Eq. 2, corresponding to an observational angle of
$\xi$ = 90$^{\circ}$ (equatorial view), is obtained for a
lightcurve amplitude:

\begin{equation}
\Delta{m} = 2.5~log\left(\frac{a}{b}\right)
\end{equation}
Since we do not know the exact spin axis orientation of each object, or the
observational angle, with Eq. 3 we can only compute a lower limit to a/b for
each object. In other words, only a lower limit to the object elongation can
be derived from the lightcurve amplitude. A lower limit to the density using
the rotational period and the lower limit on the elongation of a body can be
obtained by making use of the work of \cite{Chandrasekhar1987} for fluid
bodies. His work provides regions in the density and rotation rate space
where different Jacobi shapes of varying elongations are allowed. It also
shows that an ellipsoidal object with a/b$>$2.31 is unstable. In Table 2, we
present lower limits to the density of the bodies in this work based on the
assumptions of hydrostatic equilibrium and that the
variability is caused exclusively by shape effects. The validity of these
assumptions are discussed below. Lower limits to the densities range from 0.22~$g/cm^{3}$
for 2001~YH$_{140}$ to 2.65$\pm$0.43~$g/cm^{3}$ for 2007~UL$_{126}$. In other
words, 2001~YH$_{140}$ would have a Jacobi shape, if its density were at least
0.22~$g/cm^{3}$. The average of the lower limits to the density of our
sample is 0.90~$g/cm^{3}$. But since we have a large enough sample,
we can assume that the average viewing geometry of our bodies is
60$^{\circ}$; thus we can derive not only lower limit to the a/b ratios, but
almost true a/b ratios (in a statistically correct sense) by dividing them
by sin(60$^{\circ}$). In this way, the average density obtained from the
increased elongations would be closer to the true mean density. The value
we obtained in this case is 0.92~g/cm$^{3}$. This value should be regarded as a very rough estimation because it is
obvious that most of the lightcurves in this work are more significantly affected by albedo effects, than shape effects. A more adequate derivation of
densities is presented in \cite{Duffard2009}.

The density of an object depends on its internal composition.
However, to explain the very low densities,
$\lesssim$ 1~g/cm$^{3}$, it is helpful to consider the concept of
porosity. For example, \cite{Jewitt-Sheppard2002}, suggested that
the low density of Varuna is due to porosity. Some objects have
a higher density $\gg$ 1~g/cm$^{3}$, which suggests
that they are primarily composed of rock and ice. Objects of a high
density and large diameter might have a core of rock and a mantle of ice.
\cite{Lacerda-Jewitt2008} proposed that the high density of Haumea
is consistent with this body being the core of a
large differentiated body whose interior became exposed due to a
large collision that completely eroded its mantle.

The density of all the objects as a function of the H parameter (a
proxy for size) is shown in Fig 4. A linear fit (Dotted line)
shows almost no dependence on size. Based on a few
TNOs whose Jacobi shape is very likely, \cite{Sheppard2008}
suggested a relation between density and diameter: the largest objects
(brightest) are denser than the smallest (faintest). The fit in Fig
4 is not consistent with that idea, but one should keep in mind that most of the objects in our sample are probably MacLaurin spheroids, not Jacobi.

It is pertinent to assess whether the hydrostatic equilibrium
assumption can be applicable to the objects in our sample.
\cite{Tancredi2008} addressed the issue of the minimum
diameter needed for an object so that its mass can overcome the
rigid body forces and thus adopt a hydrostatic equilibrium shape
to become a dwarf planet, according to the 2006 IAU definition. As
mentioned by \cite{Tancredi2008}, different criteria can be used.
By integrating the hydrostatic differential equation with various
assumptions one arrives at several expressions that relate the
critical radius (R) for a self-gravitating body, the density
($\rho$), and the material strength (S). These equations can be
collectively expressed

\begin{equation}
R\rho = \sqrt{\frac{3\alpha^{2}S}{2 \pi G}}
\end{equation}
where $\alpha$ can take several values according to the different criteria
used (and ranges from $\alpha$=1 in the most simplistic case, to $\alpha$ =
$5^{{\frac{1}{2}}}$ for a spherical body in more realistic cases). We note that
a similar expression by \cite{Tancredi2008} must contain a typo because the
diameter in their equation should be radius.

One can express the size of a body using its albedo (p), absolute
magnitude $(H_{v})$ and the magnitude of the Sun $(V_{sun})$. The diameter
(D) is expressed in \cite{Russell1916} as

\begin{equation}
D=2\sqrt{\frac{2.24\times10^{16}\times10^{0.4(V_{sun}-H_{v})}}{p}}
\end{equation}
Therefore, one can express the condition for hydrostatic
equilibrium in terms of H, density, albedo and strength.

In Fig 4, we overplot the curves of density  above which
hydrostatic equilibrium is met, as a function of H. We considered
three values of material strength: 0.01, 1, and 100~MPa. We chose
two albedos values: 0.04 and 0.09. We note that Centaurs require a
much lower material strength to be in hydrostatic equilibrium
while TNOs may have more internal cohesion.

Finally, Figs. 5 and 6 are histograms of the rotation periods
based on all our data and only the TNOs presented in this
work, respectively. We assume that all the lightcurves with amplitudes below
0.15~mag are single peaked (which is almost equivalent to
assuming that their rotational variation is caused mainly by
albedo markings) and those with amplitudes larger than 0.15~mag are
double peaked. In a first step, we used this arbitrary value which had already been used by several investigators. But, to determine
if there is a more suitable value that marks the transition albedo-dominated lightcurves to shape-dominated lightcurves, we
tested with limits of 0.10~mag from 0.20~mag.

In the case of asteroids, the distribution of rotation rates is
Maxwellian \citep{Binzel1989}. A Maxwellian distribution has the
form:

\begin{equation}
f(\Omega)=\sqrt{\frac{2}{\pi}}\frac{N\Omega^{2}}{\sigma^{3}}exp\left[\frac{-\Omega^{2}}{2\sigma^{2}}\right]
\end{equation}
where $\Omega$ is the rotation rate (Cycles/day) and N the total number of objects.

A Maxwellian fit to the rotation rate distribution with the threshold of 0.1~mag gives values of 7.5~h
(3.19~cycles/day) for the whole sample and 7.3~h (3.29~cycles/day) for the TNOs alone.
A Maxwellian fit to the rotation rate distribution with the threshold of 0.15~mag infers values of 7.3~h (3.27~cycles/day) for the whole sample, 8.1~h (2.98~cycles/day)
for the TNOs alone.
A Maxwellian fit to the rotation rate distribution with the threshold of 0.20~mag gives values of 8~h
(3~cycles/day) for the whole sample and 8.1~h (2.98~cycles/day) for the TNOs alone.
With only 6 periods of Centaurs, we can't give a satisfactory study. For these objects, we find a mean rotational value of 7.3~h.
The confidence level of the chi-square test is higher for the threshold of 0.10~mag for the whole sample than for the TNOs
alone.

None of the three different fits was significantly better than the
others in terms of residuals because our sample of only  29
objects remains small for that purpose. Thus, we chose the
intermediate 0.15~mag threshold as perhaps the most adequate one,
based on our previous experience on asteroids and because a
0.15~mag variability is much easier to measure than 0.1~mag (and is
therefore a good limit in terms of instrumental requirements).
Fits to Maxwellian distributions of a larger sample
are shown in \cite{Duffard2009}, where the results of the present
work and a compilation of the scientific literature are used.

If we do not fit any distribution but just consider the mean rotation
periods of our objects, we obtain 7.5~h for the whole sample, 7.6~h
for the TNOs alone, and 7.3~h for the Centaurs. These estimates may
be more appropriate to compare with the average of
8.5~h quoted in \cite{Sheppard2008}. Our values imply a more rapid rotation than previously derived.

\section{Conclusions}

We have presented a large sample of Kuiper Belt Objects whose short-term variability has been studied in detail to increase the number of objects studied so far and try to
avoid observational biases. Amplitudes and rotation periods have been derived for
all of them with different degrees of reliability, but we have
compiled an ensemble of all of them to study the whole
population. We present therefore a homogeneous data set from which
some conclusions can be drawn. We found that the percentage of low
amplitude rotators is higher than previously thought and that in
our sample the rotation rates appear to be slightly higher (faster
objects) than previously suggested.

A simple idea investigated in detail in
\cite{Duffard2009} to explain the large abundance of small
amplitude objects might be that hydrostatic equilibrium
is applicable to the overwhelming majority of the bodies and that
the usual KBO shapes are MacLaurin spheroids which therefore do not
cause any shape induced variations (and whose variability is
caused by albedo variegations exclusively). We estimate that
0.1~mag seems to be a good measure of the typical variability
caused by albedo features.

The plots of both amplitude versus size and rotation rate versus size seem to
be compatible with the typical collisional evolution scenario in
which larger objects have been only slightly affected by
collisions, whereas the small fragments are highly collisionally
evolved bodies with usually more rapid spins of larger
amplitudes.

Based on the assumption of hydrostatic equilibrium, one can derive
densities for all the bodies and we found a possible trend of
higher densities toward higher sizes, which is a physically
plausible scenario. There appears to be a spin barrier that allows
us to obtain a density limit that is also compatible with the
average density derived based on hydrostatic equilibrium assumptions.
Nevertheless, a more appropriate derivation of mean densities is presented in \cite{Duffard2009}.

\begin{acknowledgements}

We are grateful to the Sierra Nevada Observatory, Calar Alto and
INT staffs. This research was based on data obtained at  the
Observatorio de Sierra Nevada which is  operated by the Instituto
de  Astrof\'{\i}sica de Andaluc\'{\i}a, CSIC. This research is
also based on observations collected at the Centro Astron\'omico
Hispano Alem\'an (CAHA) at Calar Alto, operated jointly by the
Max-Planck Institut f\"{u}r Astronomie and the Instituto de
Astrof\'{i}sica de Andaluc\'{i}a (CSIC). Other results were
obtained at the Isaac Newton Telescope. The Isaac Newton Telescope
is operated on the island of La Palma by the Isaac Newton Group in
the Spanish Observatorio del Roque de Los Muchachos of the
Instituto de Astrof\'{\i}sica de Canarias. This work was supported
by contracts AYA2008-06202-C03-01 and AYA2005-07808-C03-01. RD
acknowledges financial support from the MEC (contract Juan de la
Cierva).

\end{acknowledgements}

\bibliographystyle{aa}

\bibliography{biblio}

\begin{thebibliography}{50}
\expandafter\ifx\csname natexlab\endcsname\relax\def\natexlab#1{#1}\fi

\bibitem[{{Bauer} {et~al.}(2003){Bauer}, {Meech}, {Fern{\'a}ndez},
  {Pittichova}, {Hainaut}, {Boehnhardt}, \& {Delsanti}}]{Bauer2003}
{Bauer}, J.~M., {Meech}, K.~J., {Fern{\'a}ndez}, Y.~R., {et~al.} 2003, Icarus,
  166, 195

\bibitem[{{Belskaya} {et~al.}(2006){Belskaya}, {Ortiz}, {Rousselot}, {Ivanova},
  {Borisov}, {Shevchenko}, \& {Peixinho}}]{Belskaya2006}
{Belskaya}, I.~N., {Ortiz}, J.~L., {Rousselot}, P., {et~al.} 2006, Icarus, 184,
  277

\bibitem[{{Benavidez} \& {Campo Bagatin}(2009)}]{Benavidez2009}
{Benavidez}, P.~G. \& {Campo Bagatin}, A. 2009, \planss, 57, 201

\bibitem[{{Binzel} {et~al.}(1989){Binzel}, {Farinella}, {Zappala}, \&
  {Cellino}}]{Binzel1989}
{Binzel}, R.~P., {Farinella}, P., {Zappala}, V., \& {Cellino}, A. 1989, in
  Asteroids II, ed. R.~P. {Binzel}, T.~{Gehrels}, \& M.~S. {Matthews}, 416--441

\bibitem[{{Brown} {et~al.}(2009){Brown}, {Ragozzine}, {Stansberry}, \&
  {Fraser}}]{Brown2009}
{Brown}, M.~E., {Ragozzine}, D., {Stansberry}, J., \& {Fraser}, W.~C. 2009,
  ArXiv e-prints

\bibitem[{{Brown} \& {Suer}(2007)}]{Brown2007}
{Brown}, M.~E. \& {Suer}, T. 2007, \iaucirc, 8812, 1

\bibitem[{{Chandrasekhar}(1987)}]{Chandrasekhar1987}
{Chandrasekhar}, S. 1987, {Ellipsoidal figures of equilibrium} (New York :
  Dover, 1987.)

\bibitem[{{Davidsson}(1999)}]{Davidsson1999}
{Davidsson}, B.~J.~R. 1999, Icarus, 142, 525

\bibitem[{{Davidsson}(2001)}]{Davidsson2001}
{Davidsson}, B.~J.~R. 2001, Icarus, 149, 375

\bibitem[{{Davis} \& {Farinella}(1997)}]{Davies1997}
{Davis}, D.~R. \& {Farinella}, P. 1997, Icarus, 125, 50

\bibitem[{{Dotto} {et~al.}(2008){Dotto}, {Perna}, {Barucci}, {Rossi}, {de
  Bergh}, {Doressoundiram}, \& {Fornasier}}]{Dotto2008}
{Dotto}, E., {Perna}, D., {Barucci}, M.~A., {et~al.} 2008, \aap, 490, 829

\bibitem[{{Duffard} {et~al.}(2009){Duffard}, {Ortiz}, {Thirouin},
  {Santos-Sanz}, \& {Morales}}]{Duffard2009}
{Duffard}, R., {Ortiz}, J.~L., {Thirouin}, A., {Santos-Sanz}, P., \& {Morales},
  N. 2009, \aap, 505, 1283

\bibitem[{{Farnham}(2001)}]{Farnham2001}
{Farnham}, T.~L. 2001, \iaucirc, 7583, 4

\bibitem[{{Foster}(1995)}]{Foster1995}
{Foster}, G. 1995, \aj, 109, 1889

\bibitem[{{Grundy} {et~al.}(2008){Grundy}, {Noll}, {Virtanen}, {Muinonen},
  {Kern}, {Stephens}, {Stansberry}, {Levison}, \& {Spencer}}]{Grundy2008}
{Grundy}, W.~M., {Noll}, K.~S., {Virtanen}, J., {et~al.} 2008, Icarus, 197, 260

\bibitem[{{Harris} {et~al.}(1989){Harris}, {Young}, {Bowell}, {Martin},
  {Millis}, {Poutanen}, {Scaltriti}, {Zappala}, {Schober}, {Debehogne}, \&
  {Zeigler}}]{Harris1989}
{Harris}, A.~W., {Young}, J.~W., {Bowell}, E., {et~al.} 1989, Icarus, 77, 171

\bibitem[{{Holsapple}(2001)}]{Holsapple2001}
{Holsapple}, K.~A. 2001, Icarus, 154, 432

\bibitem[{{Holsapple}(2004)}]{Holsapple2004}
{Holsapple}, K.~A. 2004, Icarus, 172, 272

\bibitem[{{Jewitt} \& {Sheppard}(2002)}]{Jewitt-Sheppard2002}
{Jewitt}, D.~C. \& {Sheppard}, S.~S. 2002, \aj, 123, 2110

\bibitem[{{Lacerda} {et~al.}(2008){Lacerda}, {Jewitt}, \&
  {Peixinho}}]{Lacerda-Jewitt2008}
{Lacerda}, P., {Jewitt}, D., \& {Peixinho}, N. 2008, \aj, 135, 1749

\bibitem[{{Lacerda} \& {Luu}(2006)}]{Lacerda-Luu2006}
{Lacerda}, P. \& {Luu}, J. 2006, \aj, 131, 2314

\bibitem[{{Lagerkvist} {et~al.}(1989){Lagerkvist}, {Harris}, \&
  {Zappala}}]{Lagerkvist1989}
{Lagerkvist}, C.-I., {Harris}, A.~W., \& {Zappala}, V. 1989, in Asteroids II,
  ed. R.~P. {Binzel}, T.~{Gehrels}, \& M.~S. {Matthews}, 1162--1179

\bibitem[{{Lin} {et~al.}(2007){Lin}, {Wu}, \& {Ip}}]{Lin2007}
{Lin}, H.-W., {Wu}, Y.-L., \& {Ip}, W.-H. 2007, Advances in Space Research, 40,
  238

\bibitem[{{Lomb}(1976)}]{Lomb1976}
{Lomb}, N.~R. 1976, \apss, 39, 447

\bibitem[{{Luu} \& {Jewitt}(1998)}]{Luu-Jewitt1998}
{Luu}, J.~X. \& {Jewitt}, D.~C. 1998, \apjl, 494, L117+

\bibitem[{{Magnusson} \& {Lagerkvist}(1991)}]{Magnusson-Lagerkvist1991}
{Magnusson}, P. \& {Lagerkvist}, C.-I. 1991, \aaps, 87, 269

\bibitem[{{Morbidelli} {et~al.}(2005){Morbidelli}, {Levison}, {Tsiganis}, \&
  {Gomes}}]{Morbidelli2005}
{Morbidelli}, A., {Levison}, H.~F., {Tsiganis}, K., \& {Gomes}, R. 2005, \nat,
  435, 462

\bibitem[{{Moullet} {et~al.}(2008){Moullet}, {Lellouch}, {Doressoundiram},
  {Ortiz}, {Duffard}, {Morbidelli}, {Vernazza}, \& {Moreno}}]{Moullet2008}
{Moullet}, A., {Lellouch}, E., {Doressoundiram}, A., {et~al.} 2008, \aap, 483,
  L17

\bibitem[{{Noll} {et~al.}(2006){Noll}, {Levison}, {Stephens}, \&
  {Grundy}}]{Noll2006}
{Noll}, K.~S., {Levison}, H.~F., {Stephens}, D.~C., \& {Grundy}, W.~M. 2006,
  \iaucirc, 8751, 1

\bibitem[{{Ortiz} {et~al.}(2003{\natexlab{a}}){Ortiz}, {Guti{\'e}rrez},
  {Casanova}, \& {Sota}}]{Ortiz2003a}
{Ortiz}, J.~L., {Guti{\'e}rrez}, P.~J., {Casanova}, V., \& {Sota}, A.
  2003{\natexlab{a}}, \aap, 407, 1149

\bibitem[{{Ortiz} {et~al.}(2006){Ortiz}, {Guti\'errez}, {Santos-Sanz},
  {Casanova}, \& {Sota}}]{Ortiz2006}
{Ortiz}, J.~L., {Guti\'errez}, P.~J., {Santos-Sanz}, P., {Casanova}, V., \&
  {Sota}, A. 2006, \aap, 447, 1131

\bibitem[{{Ortiz} {et~al.}(2003{\natexlab{b}}){Ortiz}, {Guti{\'e}rrez}, {Sota},
  {Casanova}, \& {Teixeira}}]{Ortiz2003b}
{Ortiz}, J.~L., {Guti{\'e}rrez}, P.~J., {Sota}, A., {Casanova}, V., \&
  {Teixeira}, V.~R. 2003{\natexlab{b}}, \aap, 409, L13

\bibitem[{{Ortiz} {et~al.}(2007){Ortiz}, {Santos Sanz}, {Guti{\'e}rrez},
  {Duffard}, \& {Aceituno}}]{Ortiz2007}
{Ortiz}, J.~L., {Santos Sanz}, P., {Guti{\'e}rrez}, P.~J., {Duffard}, R., \&
  {Aceituno}, F.~J. 2007, \aap, 468, L13

\bibitem[{{Ortiz} {et~al.}(2004){Ortiz}, {Sota}, {Moreno}, {Lellouch}, {Biver},
  {Doressoundiram}, {Rousselot}, {Guti{\'e}rrez}, {M{\'a}rquez}, {Gonz{\'a}lez
  Delgado}, \& {Casanova}}]{Ortiz2004}
{Ortiz}, J.~L., {Sota}, A., {Moreno}, R., {et~al.} 2004, \aap, 420, 383

\bibitem[{{Pravec} \& {Harris}(2000)}]{Pravec-Harris2000}
{Pravec}, P. \& {Harris}, A.~W. 2000, Icarus, 148, 12

\bibitem[{{Press} {et~al.}(1992){Press}, {Teukolsky}, {Vetterling}, \&
  {Flannery}}]{Press1992}
{Press}, W.~H., {Teukolsky}, S.~A., {Vetterling}, W.~T., \& {Flannery}, B.~P.
  1992, {Numerical recipes in FORTRAN. The art of scientific computing}
  (Cambridge: University Press, |c1992, 2nd ed.)

\bibitem[{{Rabinowitz} {et~al.}(2006){Rabinowitz}, {Barkume}, {Brown}, {Roe},
  {Schwartz}, {Tourtellotte}, \& {Trujillo}}]{Rabinowitz2006}
{Rabinowitz}, D.~L., {Barkume}, K., {Brown}, M.~E., {et~al.} 2006, \apj, 639,
  1238

\bibitem[{{Rabinowitz} {et~al.}(2008){Rabinowitz}, {Schaefer}, {Schaefer}, \&
  {Tourtellotte}}]{Rabinowitz2008}
{Rabinowitz}, D.~L., {Schaefer}, B.~E., {Schaefer}, M., \& {Tourtellotte},
  S.~W. 2008, \aj, 136, 1502

\bibitem[{{Rabinowitz} {et~al.}(2007){Rabinowitz}, {Schaefer}, \&
  {Tourtellotte}}]{Rabinowitz2007}
{Rabinowitz}, D.~L., {Schaefer}, B.~E., \& {Tourtellotte}, S.~W. 2007, \aj,
  133, 26

\bibitem[{{Ragozzine} \& {Brown}(2007)}]{Ragozzine2007}
{Ragozzine}, D. \& {Brown}, M.~E. 2007, \aj, 134, 2160

\bibitem[{{Romanishin} \& {Tegler}(1999)}]{Romanishin-Tegler1999}
{Romanishin}, W. \& {Tegler}, S.~C. 1999, \nat, 398, 129

\bibitem[{{Russell}(1916)}]{Russell1916}
{Russell}, H.~N. 1916, \apj, 43, 173

\bibitem[{{Sheppard}(2007)}]{Sheppard2007}
{Sheppard}, S.~S. 2007, \aj, 134, 787

\bibitem[{{Sheppard} \& {Jewitt}(2002)}]{Sheppard-Jewitt2002}
{Sheppard}, S.~S. \& {Jewitt}, D.~C. 2002, in ESA Special Publication, Vol.
  500, Asteroids, Comets, and Meteors: ACM 2002, ed. B.~{Warmbein}, 21--24

\bibitem[{{Sheppard} \& {Jewitt}(2003)}]{Sheppard-Jewitt2003}
{Sheppard}, S.~S. \& {Jewitt}, D.~C. 2003, Earth Moon and Planets, 92, 207

\bibitem[{{Sheppard} {et~al.}(2008){Sheppard}, {Lacerda}, \&
  {Ortiz}}]{Sheppard2008}
{Sheppard}, S.~S., {Lacerda}, P., \& {Ortiz}, J.~L. 2008, {Photometric
  Lightcurves of Transneptunian Objects and Centaurs: Rotations, Shapes, and
  Densities} (The Solar System Beyond Neptune), 129--142

\bibitem[{{Stetson}(1987)}]{Stetson1987}
{Stetson}, P.~B. 1987, \pasp, 99, 191

\bibitem[{{Tancredi} \& {Favre}(2008)}]{Tancredi2008}
{Tancredi}, G. \& {Favre}, S. 2008, Icarus, 195, 851

\bibitem[{{Tiscareno} \& {Malhotra}(2003)}]{Tiscareno2003}
{Tiscareno}, M.~S. \& {Malhotra}, R. 2003, \aj, 126, 3122

\bibitem[{{Trilling} \& {Bernstein}(2006)}]{Trilling-Bernstein2006}
{Trilling}, D.~E. \& {Bernstein}, G.~M. 2006, \aj, 131, 1149

\end{thebibliography}


\begin{figure*}
   \centering
   \includegraphics[angle=0, width=15cm]{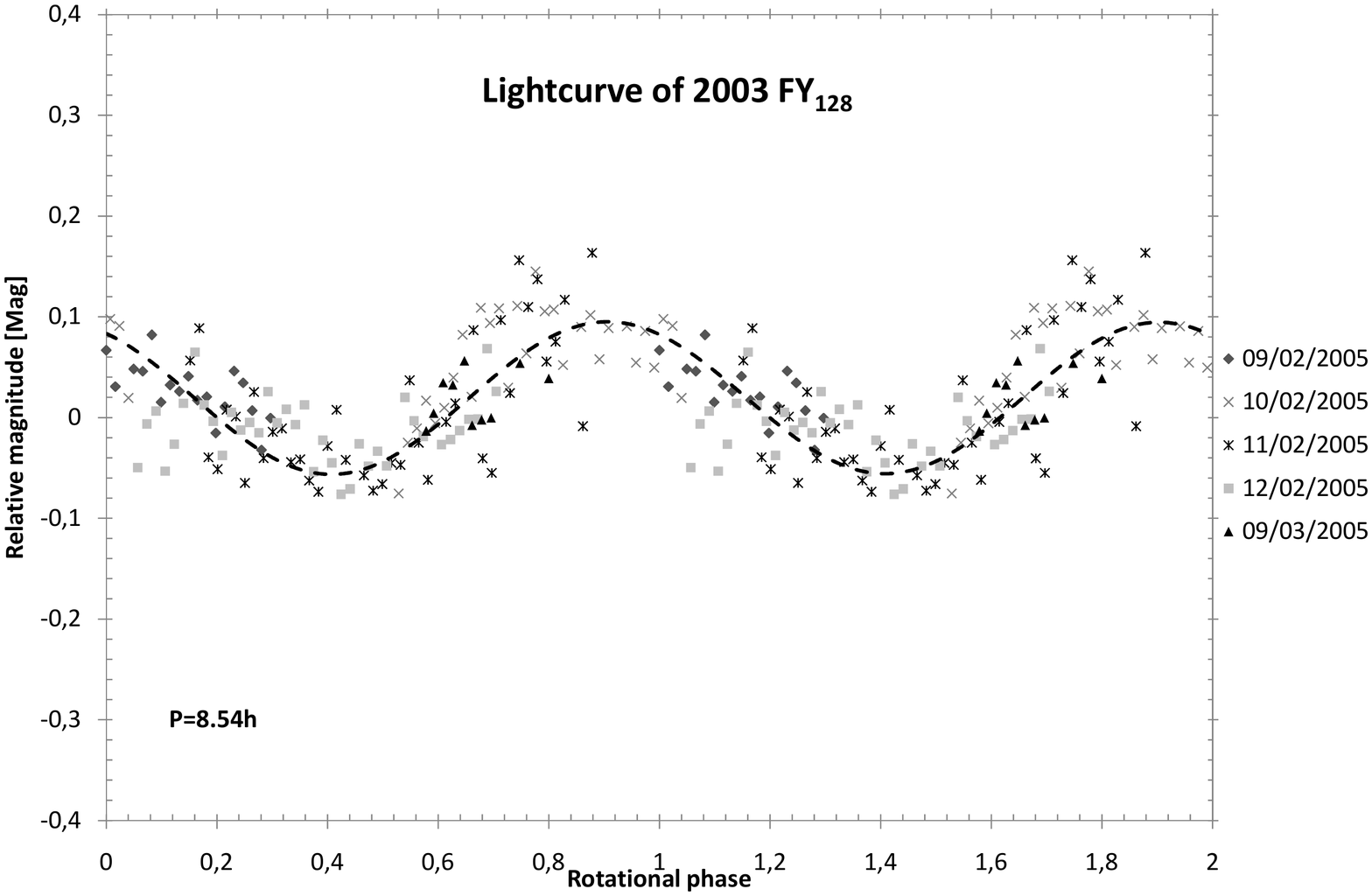}
   \includegraphics[angle=0, width=15cm]{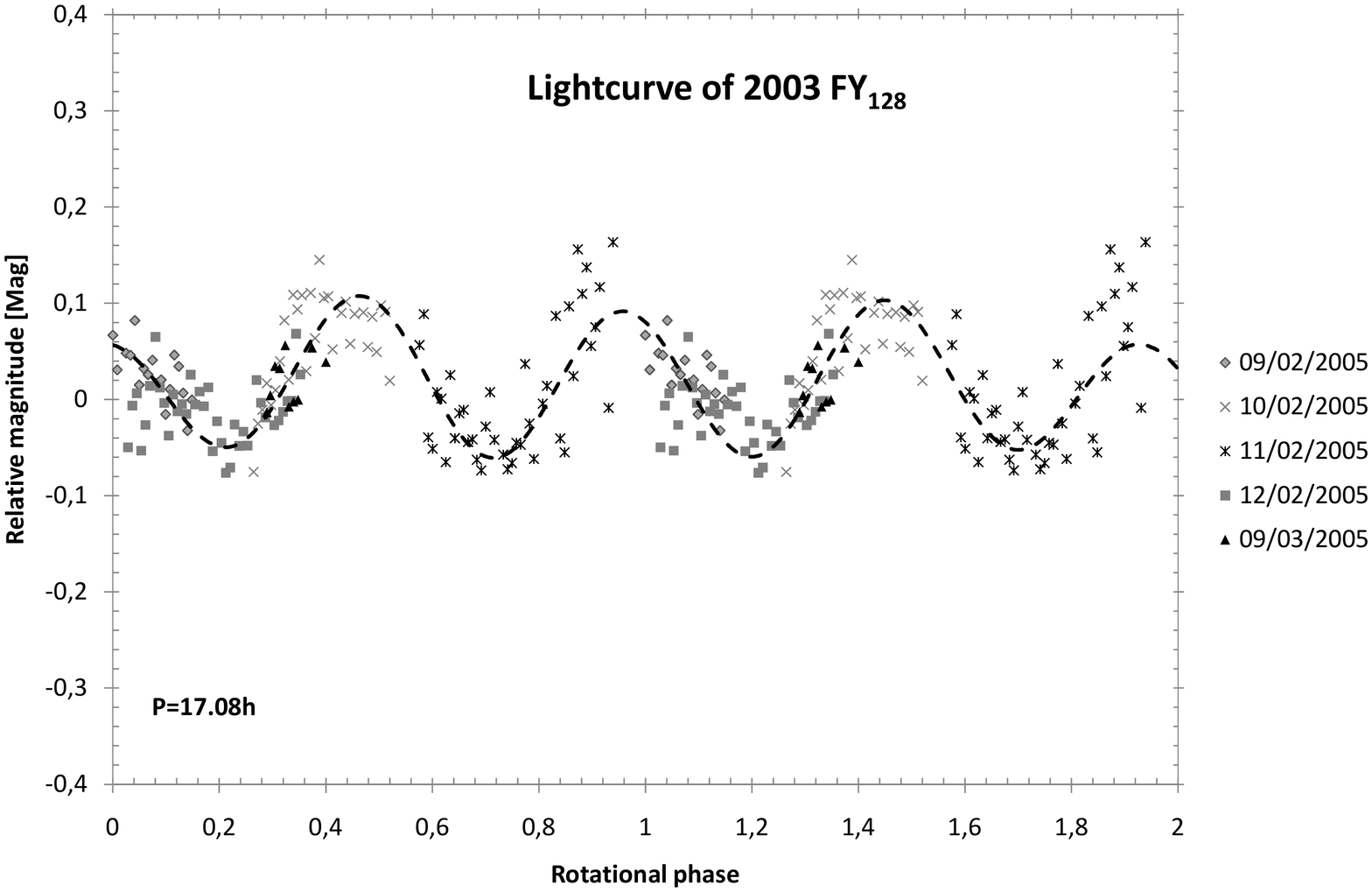}
\caption{Rotational phase curve for 2003~FY$_{128}$ obtained by using a spin
period of 8.54~h (upper plot) and a spin period of 17.08~h (lower plot). The
dash line is a Fourier series fit of the photometric data. Different symbols
correspond to different dates.}
\end{figure*}

\clearpage

\begin{figure*}
   \centering
   \includegraphics[angle=0, width=15cm]{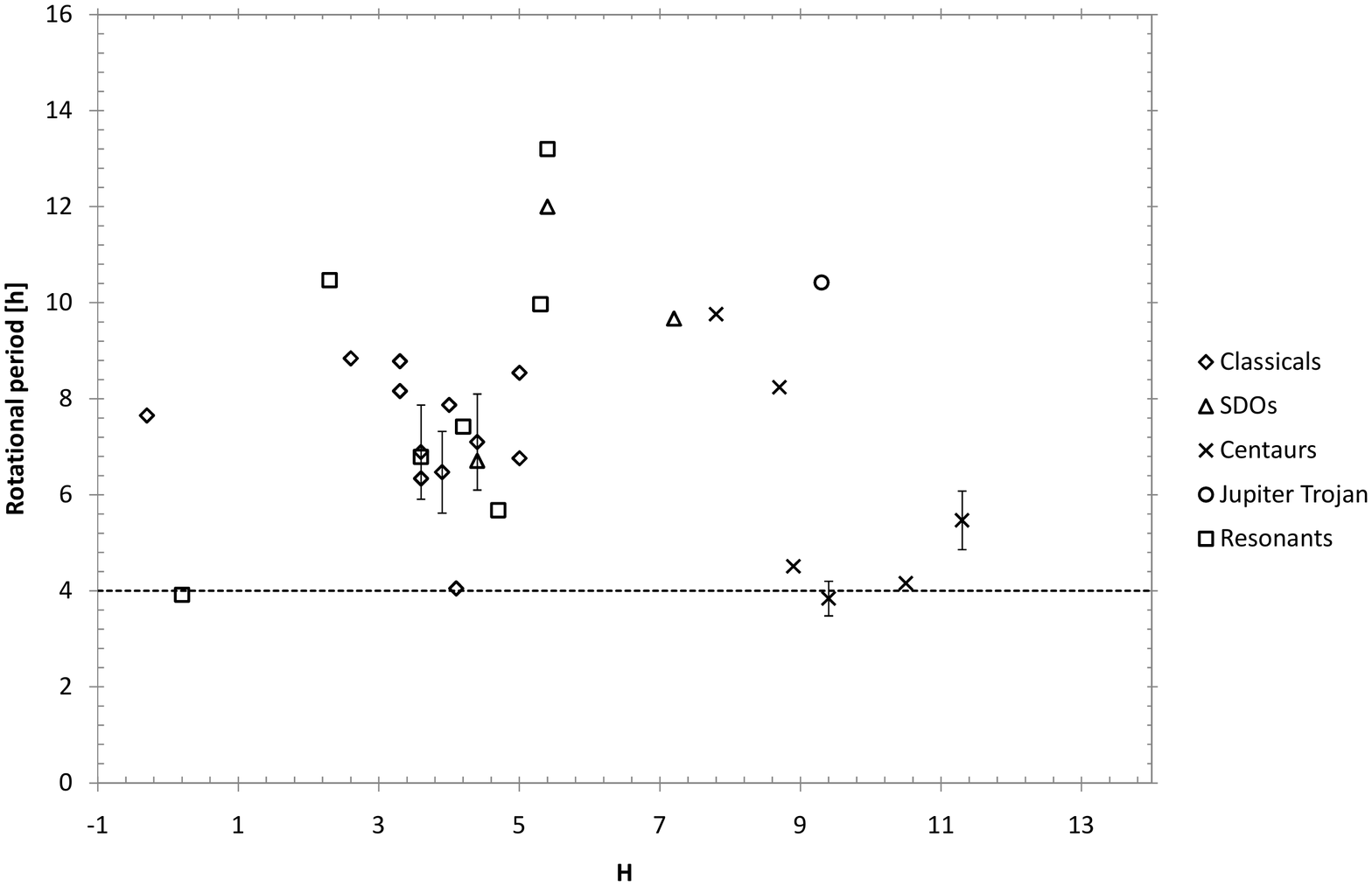}
\caption{Rotational period versus H for all objects presented in this
work. Different symbols correspond to different object
classification. Dashed line defines a spin barrier.}
\end{figure*}

\clearpage

\begin{figure*}
   \centering
   \includegraphics[angle=0, width=15cm]{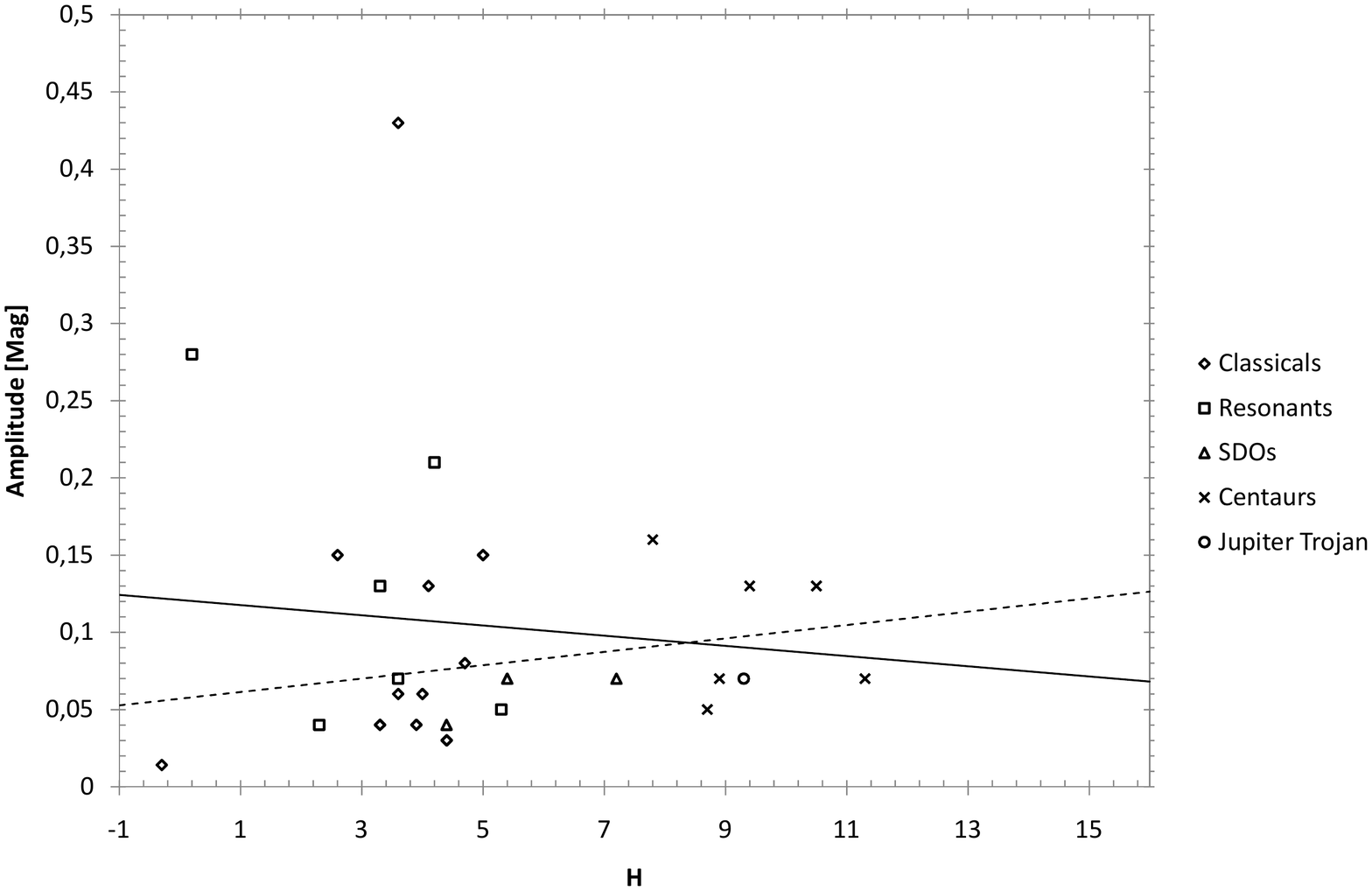}
\caption{Amplitude versus H for all objects presented in this work.
Different symbols correspond to different object classifications.
Thick line is a linear fit of all our sample of data. Dash line is
a linear fit to all our sample of data except Varuna, Haumea and
2003~VS$_{2}$ which exhibit a peak-to-peak amplitude $>$0.20~mag. }
\end{figure*}

\clearpage

\begin{figure*}
   \centering
   \includegraphics[angle=0, width=15cm]{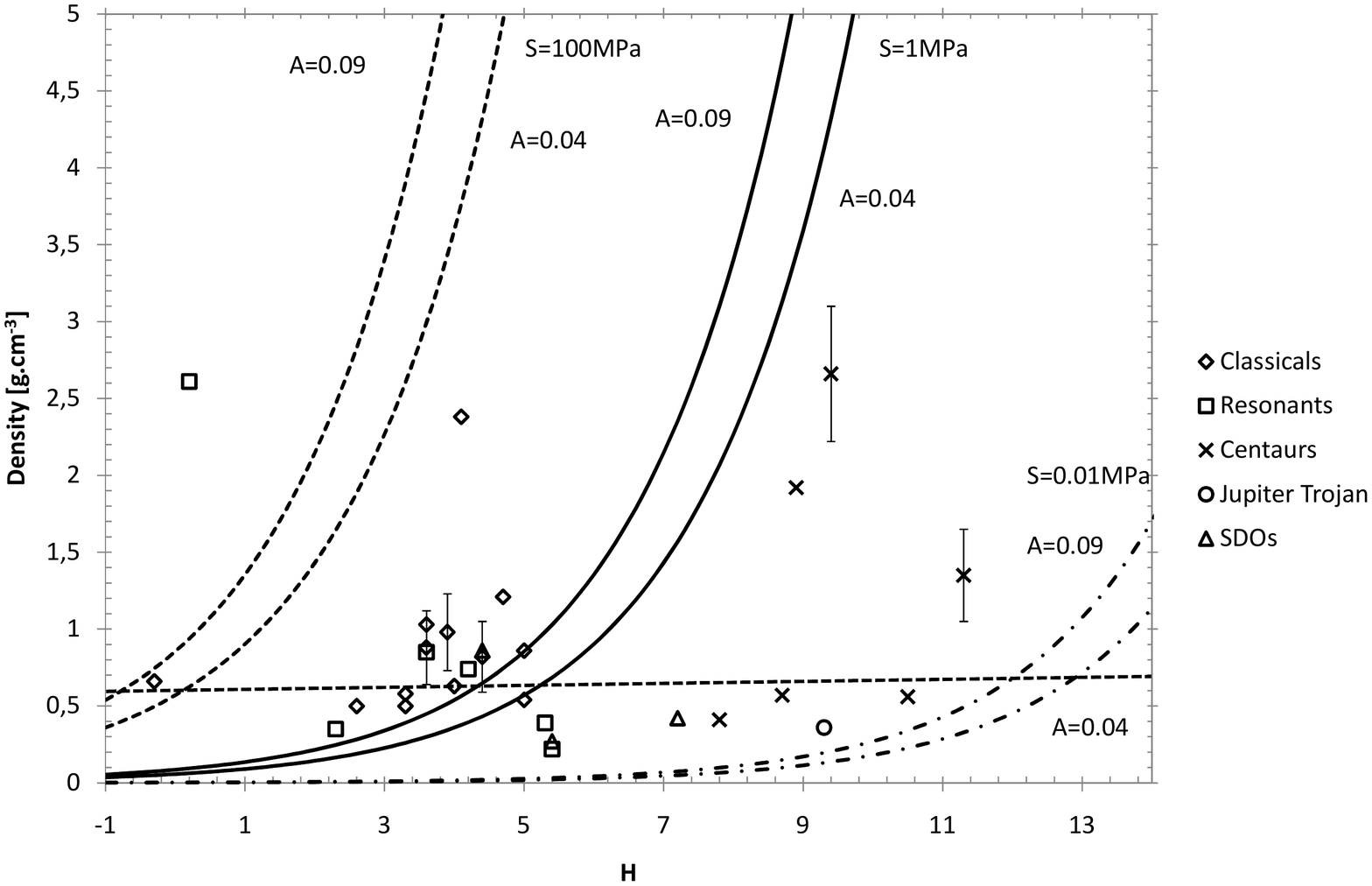}
\caption{Density versus H for all objects presented in this work. Different
symbols correspond to different object classification. Dotted line is a linear
fit. See text for definitions of the meaning of remaning.}
\end{figure*}

\clearpage

\begin{figure*}
   \centering
   \includegraphics[angle=0, width=15cm]{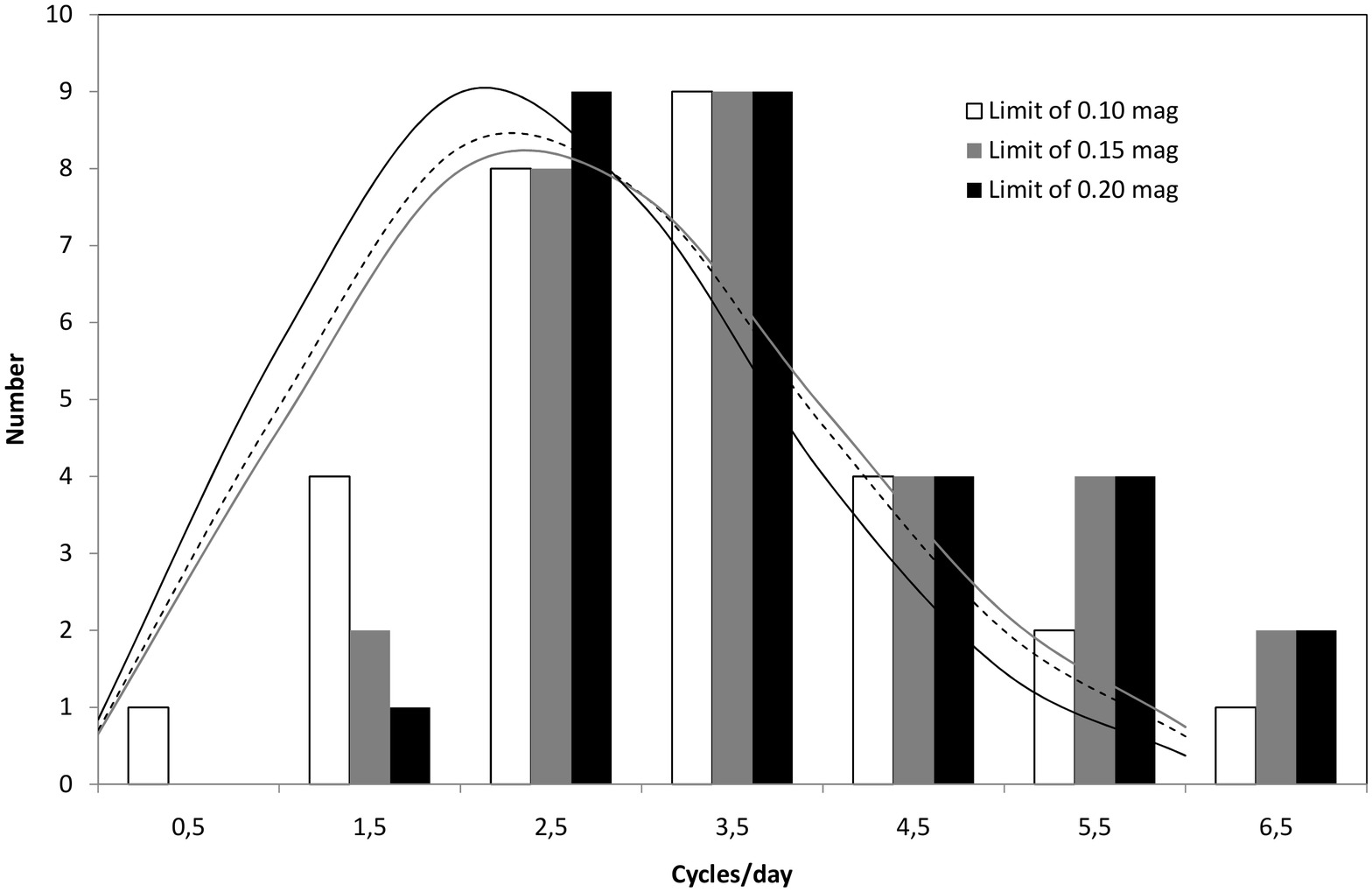}
\caption{Histograms of all our sample of data. We assume that all
the lightcurves with amplitudes below an arbitrary limit are
single peaked and those with amplitudes larger than this limit are
double peaked. White data correspond to a limit of 0.10~mag, grey
data correspond to a limit of 0.15~mag and black data correspond to
a limit of 0.20~mag. Dash line is the Maxwellian fit to the
distribution with 0.10~mag limit. Grey line is the Maxwellian
fit to the distribution with 0.15~mag limit. Black line is
the Maxwellian fit to the distribution with 0.20~mag limit.}
\end{figure*}

\clearpage

\begin{figure*}
   \centering
   \includegraphics[angle=0, width=15cm]{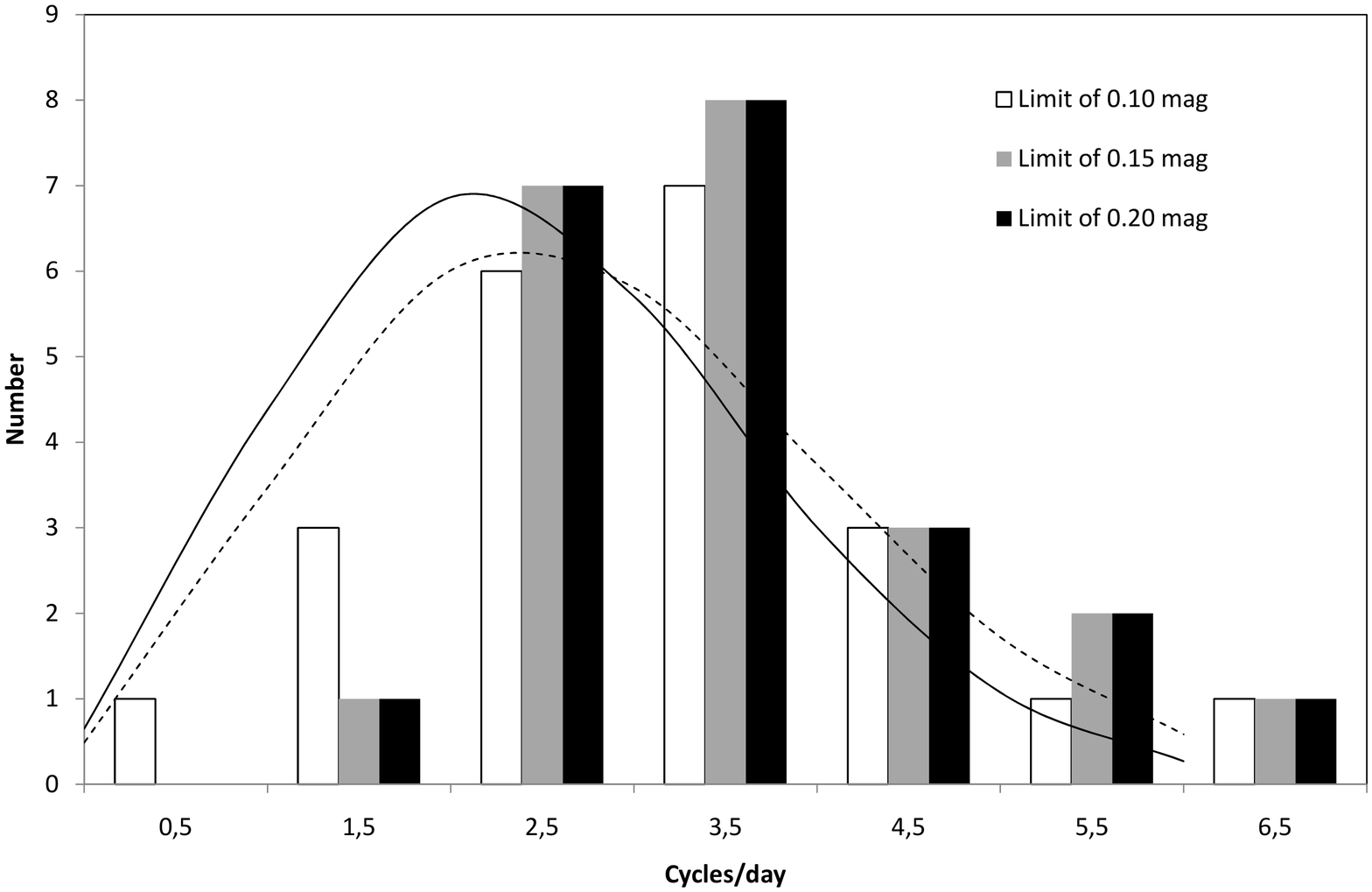}
\caption{Histograms of all TNOs presented in this work (our sample
without the Centaurs). We assume that all the lightcurves with
amplitudes below an arbitrary limit are single peaked and those
with amplitude larger than this limit are double peaked. White
data correspond to a limit of 0.10~mag, grey data correspond to a
limit of 0.15~mag and black data correspond to a limit of 0.20~mag.
Dashed line is the Maxwellian fit to the distribution with
0.10~mag limit. Grey line is the Maxwellian fit to the
distribution with 0.15~mag limit. Black line is the Maxwellian
fit to the distribution with 0.20~mag limit. Grey line
falls on top of black line. }
\end{figure*}


\longtab{1}{
\begin{longtable}{lccccccc}
\caption{\label{tabdatgeo1}Dates, geometric and photometric data
of the observations}\\
\hline\hline Object  & Date & \# Images & r$_\mathrm{h}$[AU] &
$\Delta$ [AU] & $\alpha$[deg] & Filter & Telescope \\ \hline
\endfirsthead
\caption{continued.}\\
\hline\hline Object  & Date & \# Images & r$_\mathrm{h}$[AU] &
$\Delta$ [AU] & $\alpha$[deg] & Filter & Telescope \\ \hline
\endhead
\hline
\endfoot
(15874) 1996~TL$_{66}$
          & 15/12/2004  & 16 & 35.141 & 34.344 & 0.95 &  & OSN \\
          & 16/12/2004  & 16 & 35.141 & 34.354 & 0.98 &  & OSN \\
          & 17/12/2004  & 30 & 35.141 & 34.365 & 1.00 &  & OSN \\
          & 18/12/2004  & 14 & 35.141 & 34.377 & 1.02 &  & OSN \\
(52872) 1998~SG$_{35}$
          & 05/12/2007  & 26 & 5.804 & 5.621 & 9.72 &  & OSN \\
          & 06/12/2007  & 33 & 5.804 & 5.605 & 9.70 &  & OSN \\
          & 07/12/2007  & 19 & 5.804 & 5.590 & 9.68 &  & OSN \\
          & 08/12/2007  & 14 & 5.804 & 5.574 & 9.66 &  & OSN \\
          & 10/12/2007  & 40 & 5.803 & 5.542 & 9.60 &  & OSN \\
          & 11/12/2007  & 33 & 5.803 & 5.526 & 9.57 &  & OSN \\
          & 12/12/2007  & 35 & 5.803 & 5.511 & 9.53 &  & OSN \\
          & 13/12/2007  & 33 & 5.803 & 5.495 & 9.50 &  & OSN \\
          & 14/12/2007  & 38 & 5.803 & 5.479 & 9.46 &  & OSN \\
          & 15/12/2007  & 38 & 5.802 & 5.464 & 9.41 &  & OSN \\
(12929) 1999~TZ$_{1}$
          & 23/02/2007  & 10 & 5.317 & 4.732 & 9.12 & R & OSN \\
          & 25/02/2007  & 10 & 5.317 & 4.704 & 8.91 & R & OSN \\
          & 09/03/2007  & 20 & 5.314 & 4.552 & 7.42 & R & OSN \\
          & 10/03/2007  & 30 & 5.314 & 4.540 & 7.27 & R & OSN \\
          & 11/03/2007  & 29 & 5.313 & 4.529 & 7.13 & R & OSN \\
          & 12/03/2007  & 30 & 5.313 & 4.518 & 6.98 & R & OSN \\
(20000) 2000~WR$_{106}$
       & 05/01/2005 & 22 & 43.248 & 42.266 & 0.06 & R & OSN \\
       & 07/01/2005 & 13 & 43.249 & 42.267 & 0.09 & R & OSN \\
       & 31/01/2005 & 27 & 43.252 & 42.378 & 0.61 & R & OSN \\
       & 01/02/2005 & 5  & 43.252 & 42.388 & 0.63 & R & OSN \\
       & 09/02/2005 & 10 & 43.253 & 42.465 & 0.79 & R & OSN \\
       & 10/02/2005 & 11 & 43.253 & 42.474 & 0.81 & R & OSN \\
(126154) 2001~YH$_{140}$
          & 15/12/2004 & 7 & 36.436 & 35.569 & 0.74 &  & OSN \\
          & 16/12/2004 &10 & 36.436 & 35.561 & 0.71 &  & OSN \\
          & 17/12/2004 &12 & 36.436 & 35.554 & 0.69 &  & OSN \\
          & 18/12/2004 &6  & 36.436 & 35.546 & 0.67 &  & OSN \\
          & 19/12/2004 &10 & 36.436 & 35.539 & 0.64 &  & OSN \\
(55565) 2002~AW$_{197}$
          & 01/02/2003  & 100& 47.272 & 46.295 & 0.16 &  & OSN \\
          & 02/02/2003  & 66 & 47.272 & 46.295 & 0.15 &  & OSN \\
          & 19/01/2004  & 20 & 47.158 & 46.220 & 0.36 &  & OSN \\
          & 21/01/2004  & 50 & 47.158 & 46.211 & 0.33 &  & OSN \\
          & 22/01/2004  & 30 & 47.158 & 46.206 & 0.31 &  & OSN \\
          & 23/01/2004  & 45 & 47.157 & 46.202 & 0.29 &  & OSN \\
          & 24/01/2004  & 30 & 47.157 & 46.198 & 0.27 &  & OSN \\
          & 25/01/2004  & 30 & 47.157 & 46.195 & 0.26 &  & OSN \\
(42355) 2002~CR$_{46}$
          & 28/01/2003  & 109& 17.892 & 16.909 & 0.18 &  & OSN \\
          & 02/02/2003  & 69 & 17.889 & 16.905 & 0.16 &  & OSN \\
          & 04/03/2003  & 91 & 17.872 & 17.039 & 1.77 &  & OSN \\
          & 06/03/2003  & 87 & 17.870 & 17.057 & 1.87 &  & OSN \\
          & 09/03/2003  & 51 & 17.869 & 17.086 & 2.01 &  & OSN \\
(55576) 2002~GB$_{10}$
          & 08/03/2003  & 67 & 15.188 & 14.327 &1.92&  & OSN \\
          & 09/03/2003  & 64 & 15.188 & 14.319 &1.87&  & OSN \\
(50000) 2002~LM$_{60}$
          & 21/05/2003 & 30 & 43.407 & 42.421 & 0.31&   & OSN \\
          & 22/05/2003 & 77 & 43.407 & 42.418 & 0.29&   & OSN \\
          & 23/05/2003 & 98 & 43.406 & 42.415 & 0.27&   & OSN \\
          & 17/06/2003 & 18 & 43.404 & 42.431 & 0.39&   & OSN \\
          & 18/06/2003 & 38 & 43.404 & 42.434 & 0.41&   & OSN \\
          & 19/06/2003 & 62 & 43.404 & 42.439 & 0.42&   & OSN \\
          & 20/06/2003 & 65 & 43.404 & 42.444 & 0.44&   & OSN \\
          & 21/06/2003 & 45 & 43.404 & 42.449 & 0.46&   & OSN \\
          & 22/06/2003 & 12 & 43.404 & 42.455 & 0.49&   & OSN \\
(55636) 2002~TX$_{300}$
          & 07/08/2003 & 127  & 40.825 & 40.303 & 1.23&   & OSN \\
          & 08/08/2003 & 177  & 40.825 & 40.291 & 1.22 &  & OSN \\
          & 09/08/2003 & 173  & 40.825 & 40.278 & 1.20 &  & OSN \\
(55638) 2002~VE$_{95}$
          & 19/01/2004 & 10 & 28.015 & 27.650 & 1.88 &  & OSN \\
          & 14/12/2004 & 5  & 28.049 & 27.203 &1.04 &  & OSN \\
          & 15/12/2004 & 10 & 28.049 & 27.211 &1.07 &  & OSN \\
          & 16/12/2004 & 15 & 28.050 & 27.219 & 1.09 &   & OSN \\
          & 17/12/2004 & 18 & 28.050 & 27.230 &1.13 &  & OSN \\
          & 18/12/2004 & 5  & 28.050 & 27.238 & 1.15&   & OSN \\
          & 19/12/2004 & 50 & 28.050 & 27.248 & 1.18&   & OSN \\
(208996) 2003~AZ$_{84}$
           & 21/01/2004 & 15& 45.829 & 44.881 & 0.33&  & OSN \\
           & 23/01/2004 & 15& 45.828 & 44.889 & 0.37&  & OSN \\
           & 24/01/2004 & 21& 45.828 & 44.893 & 0.39&  & OSN \\
           & 25/01/2004 & 10& 45.828 & 44.898 & 0.41&  & OSN \\
           & 14/12/2004 & 5 & 45.765 & 44.890 & 0.57&  & OSN \\
           & 15/12/2004 & 19& 45.765 & 44.875 & 0.55&  & OSN \\
           & 16/12/2004 & 20& 45.765 & 44.875 & 0.53&  & OSN \\
           & 17/12/2004 & 25& 45.764 & 44.867 & 0.51&  & OSN \\
           & 18/12/2004 & 4 & 45.764 & 44.860 & 0.49&  & OSN \\
           & 19/12/2004 & 20& 45.764 & 44.855 & 0.47&  & OSN \\
(120061) 2003~CO$_{1}$
        & 19/01/2004   & 5 & 11.445  & 10.806 & 3.85&  & OSN \\
        & 21/01/2004   & 15& 11.443  & 10.780 & 3.75&  & OSN \\
        & 22/01/2004   & 20& 11.442  & 10.768 & 3.70&  & OSN \\
        & 23/01/2004   & 13& 11.441  & 10.756 & 3.65&  & OSN \\
        & 24/01/2004   & 30& 11.440  & 10.745 & 3.60&  & OSN \\
        & 25/01/2004   & 23& 11.438  & 10.732 & 3.55&  & OSN \\
        & 19/04/2004   & 71& 11.354  & 10.694 & 3.94&  & OSN \\
        & 23/04/2004   & 52& 11.350  & 10.737 & 4.14&  & OSN \\
        & 25/04/2004   & 60& 11.349  & 10.759 & 4.23&  & OSN \\
        & 26/04/2004   & 53& 11.348  & 10.771 & 4.28&  & OSN \\
        & 27/04/2004   & 10& 11.347  & 10.782 & 4.32&  & OSN \\
(136108) 2003~EL$_{61}$
          & 12/01/2007 & 6 & 51.175  & 51.059 & 1.09 & R & Calar Alto\\
          & 13/01/2007 & 7 & 51.175  & 51.043 & 1.09 & R & Calar Alto\\
          & 14/01/2007 & 8 & 51.175  & 51.028& 1.09 & R & Calar Alto\\
          & 15/01/2007 & 11 & 51.175  & 51.013 & 1.09& R  & Calar Alto\\
          & 16/01/2007 & 4 & 51.175  & 50.997 & 1.08 & R & Calar Alto\\
          & 17/01/2007 & 4 & 51.175  & 50.981 & 1.08 & R & Calar Alto\\
(120132) 2003~FY$_{128}$
           & 09/02/2005 & 19& 38.063 & 37.366 & 1.06&  & OSN \\
           & 10/02/2005 & 28& 38.063 & 37.355 & 1.04&  & OSN \\
           & 11/02/2005 & 44& 38.063 & 37.344 & 1.03&  & OSN \\
           & 12/02/2005 & 44& 38.064 & 37.332 & 1.01&  & OSN \\
           & 09/03/2005 & 13& 38.071 & 37.123 & 0.45&  & OSN \\
(174567) 2003~MW$_{12}$
           & 28/05/2006 & 14&48.188& 47.233 & 0.41&  & OSN \\
           & 05/06/2006 & 11& 48.186 &47.241&0.44 &  & OSN \\
           & 06/06/2006 & 29&48.185& 47.242 & 0.45&  & OSN \\
           & 07/06/2006 & 22&48.185& 47.245 & 0.46&  & OSN \\
           & 10/06/2006 & 20 &48.184& 47.254 & 0.49&  & OSN \\
           & 23/06/2006 & 5& 48.180 & 47.320 & 0.65&  & OSN \\
           & 24/06/2006 & 10&  48.180& 47.326 & 0.66&  & OSN \\
           & 12/04/2008 & 10& 47.968 & 47.305 & 0.90&  & OSN \\
           & 13/04/2008 & 17& 47.967 & 47.293 & 0.89&  & OSN \\
           & 14/04/2008 & 10& 47.967 & 47.280 & 0.88&  & OSN \\
           & 26/04/2008 & 36& 47.963 & 47.157 & 0.73&  & OSN \\
           & 27/04/2008 & 27& 47.963 & 47.148 & 0.71&   & OSN \\
(120178) 2003~OP$_{32}$
          & 05/08/2005 & 15 & 41.058 & 40.111 & 0.51&  & OSN \\
          & 06/08/2005 & 10 & 41.059 & 40.107 & 0.50&  & OSN \\
          & 07/08/2005 & 15 & 41.059 & 40.105 & 0.49&  & OSN \\
          & 10/08/2005 & 15 & 41.060 & 40.102 & 0.47&  & OSN \\
          & 03/10/2005 & 10 & 41.074 & 40.439 & 1.09&  & OSN \\
          & 04/10/2005 & 21 & 41.074 & 40.451 & 1.10&  & OSN \\
          & 05/10/2005 & 24 & 41.074 & 40.464 & 1.11&  & OSN \\
          & 15/09/2007 & 10 & 41.264 & 40.410 & 0.75&  & OSN \\
          & 16/09/2007 & 12 & 41.264 & 40.417 & 0.76&  & OSN \\
          & 17/09/2007 & 10 & 41.265 & 40.426 & 0.78&  & OSN \\
(84922) 2003~VS$_{2}$
           & 22/12/2003 & 34& 36.431 & 35.655 & 0.96&  & OSN \\
           & 26/12/2003 & 21& 36.431 & 35.695 & 1.04&  & OSN \\
           & 28/12/2003 & 26& 36.431 & 35.719 & 1.08&  & OSN \\
           & 04/01/2004 & 109&36.431 & 35.803 & 1.20&  & OSN \\
           & 19/01/2004 & 19 &36.431 & 36.015 & 1.41&  & OSN \\
           & 20/01/2004 & 30 &36.431 & 36.030 & 1.42&  & OSN \\
           & 21/01/2004 & 40 &36.431 & 36.046 & 1.43&  & OSN \\
           & 22/01/2004 & 50 &36.431 & 36.062 & 1.44&  & OSN \\
(136204) 2003~WL$_{7}$
           & 05/12/2007 & 51 &15.201 & 14.300 & 1.55&  & OSN \\
           & 06/12/2007 & 32 &15.201 & 14.307 & 1.61&  & OSN \\
           & 07/12/2007 & 20 &15.200 & 14.313 & 1.67&  & OSN \\
           & 08/12/2007 & 40 &15.200 & 14.320 & 1.72&  & OSN \\
           & 10/12/2007 & 35 &15.199 & 14.343 & 1.89&  & OSN \\
           & 11/12/2007 & 44 &15.198 & 14.351 & 1.95&  & OSN \\
           & 13/12/2007 & 40 &15.198 & 14.360 & 2.01&  & OSN \\
           & 14/12/2007 & 41 &15.198 & 14.368 & 2.06&  & OSN \\
(90482) 2004~DW
           & 08/03/2004 & 34 &47.612 & 46.746 & 0.59& R & OSN \\
           & 09/03/2004 & 24 &47.612 & 46.754 & 0.61& R & OSN \\
           & 10/03/2004 & 32 &47.612 & 46.761 & 0.62& R & OSN \\
           & 11/03/2004 & 16 &47.613 & 46.767 & 0.63& R & OSN \\
           & 23/03/2004 & 23 &47.615 & 46.875 & 0.81& R & OSN \\
           & 22/04/2004 & 39 &47.619 & 47.268 & 1.14& R & OSN \\
           & 23/04/2004 & 53 &47.620 & 47.283 & 1.15& R & OSN \\
           & 25/04/2004 & 48 &47.620 & 47.314 & 1.16& R & OSN \\
           & 26/04/2004 & 42 &47.620 & 47.330 & 1.16& R & OSN \\
           & 27/04/2004 & 37 &47.620 & 47.345 & 1.17& R & OSN \\
(120347) 2004~SB$_{60}$
           & 05/08/2005 & 15 & 43.702 & 42.908 & 0.83&   & OSN\\
           & 06/08/2005 & 15 & 43.702 & 42.900 & 0.82&   & OSN \\
           & 07/08/2005 & 10 & 43.702 & 42.891 & 0.81&   & OSN \\
           & 10/08/2005 & 15 & 43.703 & 42.870 & 0.76&   & OSN \\
           & 03/08/2008 & 16 & 43.979 & 43.240 & 0.91&   & OSN \\
           & 04/08/2008 & 15 & 43.979 & 43.231 & 0.90&   & OSN \\
           & 06/08/2008 & 52 & 43.980 & 43.214 & 0.87&   & OSN \\
           & 07/08/2008 & 26 & 43.980 & 43.205 & 0.86&   & OSN \\
(144897) 2004~UX$_{10}$
           & 14/09/2007 & 10 &38.824 & 38.014 & 0.89&  & OSN \\
           & 17/09/2007 & 12 &38.825 & 37.986 & 0.83&  & OSN \\
           & 30/11/2007 & 52 &38.834 & 38.102 & 0.99&  & OSN \\
2005~CB$_{79}$
           & 06/01/2008 & 22 &40.173 & 39.337 & 0.75&  & Calar Alto \\
           & 07/01/2008 & 15 &40.173 & 39.328 & 0.73&  & Calar Alto \\
           & 01/05/2008 & 14 &40.133 & 40.076 & 1.44&  & OSN \\
           & 04/05/2008 & 18 &40.132 & 40.125 & 1.44&  & OSN \\
           & 26/12/2008 & 38 &40.051 & 39.334 & 0.97&  & OSN \\
(136472) 2005~FY$_{9}$
           & 01/03/2006 & 21 &51.926 & 51.075 & 0.57& R & OSN \\
           & 02/03/2006 & 9 &51.926 & 51.073 & 0.56& R & OSN \\
           & 07/04/2006 &145 &51.932 & 51.150 & 0.69& R & OSN \\
           & 08/04/2006 & 23 &51.932 & 51.157 & 0.70& R & OSN \\
           & 10/04/2006 & 84 &51.933 & 51.171 & 0.72& R & OSN \\
           & 12/04/2006 & 55 &51.933 & 51.187 & 0.74& R & OSN \\
           & 27/05/2006 & 15 &51.941 & 51.715 & 1.09& R & OSN \\
           & 28/05/2006 & 20 &51.941 & 51.728 & 1.10& R & OSN \\
           & 29/05/2006 & 5  &51.941 & 51.744 & 1.10& R & OSN \\
           & 05/06/2006 & 5  &51.942 & 51.846 & 1.12& R & OSN \\
           & 06/06/2006 & 10 &51.942 & 51.863 & 1.12& R & OSN \\
           & 07/06/2006 & 35 &51.942 & 51.875 & 1.12& R & OSN \\
           & 10/06/2006 & 10 &51.943 & 51.922 & 1.12& R & OSN \\
           & 14/12/2006 & 31 &51.973 & 51.974 & 1.08& R & OSN \\
           & 15/12/2006 & 36 &51.973 & 51.960 & 1.08& R & OSN \\
           & 16/12/2006 & 30 &51.973 & 51.945 & 1.08& R & OSN \\
           & 17/12/2006 & 18 &51.974 & 51.930 & 1.08& R & OSN \\
           & 18/12/2006 & 5  &51.974 & 51.915 & 1.08& R & OSN \\
           & 11/01/2007 &  9 &51.978 & 51.569 & 0.99& R & Calar Alto \\
           & 12/01/2007 &  10&51.978 & 51.557 & 0.98& R & Calar Alto \\
           & 13/01/2007 &  7 &51.978 & 51.544 & 0.98& R & Calar Alto \\
           & 14/01/2007 & 9  &51.978 & 51.531 & 0.97& R & Calar Alto \\
           & 15/01/2007 & 4  &51.978 & 51.518 & 0.96& R & Calar Alto \\
           & 16/01/2007 & 5  &51.979 & 51.505 & 0.95& R & Calar Alto \\
           & 09/03/2007 & 10 &51.987 & 51.122 & 0.54& R & OSN \\
           & 10/03/2007 & 20 &51.987 & 51.121 & 0.54& R & OSN \\
           & 11/03/2007 & 32 &51.987 & 51.121 & 0.54& R & OSN \\
           & 12/03/2007 & 25 &51.987 & 51.120 & 0.54& R & OSN \\
(145451) 2005~RM$_{43}$
           & 13/10/2006 & 19 &35.139 & 34.321 & 0.94&  & OSN \\
           & 14/10/2006 & 12 &35.139 & 34.313 & 0.92&  & OSN \\
           & 15/12/2006 & 18 &35.146 & 34.356 & 0.97&  & OSN \\
           & 17/12/2006 & 12 &35.146 & 34.375 & 1.01&  & OSN \\
           & 18/12/2006 & 27 &35.146 & 34.385 & 1.03&  & OSN \\
           & 11/01/2007 & 4 & 35.149 & 34.687 & 1.43&   & Calar Alto \\
           & 12/01/2007 & 5 & 35.149 & 34.701 & 1.44&   & Calar Alto \\
           & 13/01/2007 & 5 & 35.149 & 34.716 & 1.45&   & Calar Alto \\
           & 14/01/2007 & 8 & 35.149 & 34.732 & 1.46&   & Calar Alto \\
           & 15/01/2007 & 3 & 35.149 & 34.745 & 1.47&   & Calar Alto \\
(145452) 2005~RN$_{43}$
           & 22/10/2006 & 6 & 40.723  & 40.266 & 1.25 & R & INT \\
           & 14/09/2007 & 7 & 40.714 & 39.808 & 0.62 &  & OSN \\
           & 16/09/2007 & 10 & 40.714 & 39.821 & 0.66&   & OSN \\
           & 17/09/2007 & 10 & 40.714 & 39.829 & 0.68&   & OSN \\
           & 19/09/2007 & 6  & 40.714 & 39.844 & 0.71&   & OSN \\
           & 03/08/2008 & 15 & 40.706 & 39.766 & 0.55&   & OSN \\
           & 04/08/2008 & 15 & 40.706 & 39.761 & 0.53&   & OSN \\
           & 05/08/2008 & 30 & 40.706 & 39.756 & 0.51&   & OSN \\
           & 07/08/2008 & 25 & 40.706 & 39.747 & 0.47&   & OSN \\
           & 08/08/2008 & 37 & 40.706 & 39.743 & 0.45&   & OSN \\
(145453) 2005~RR$_{43}$
           & 22/10/2006 & 10 &38.410  &37.527   & 0.69& R  & INT \\
           & 23/10/2006 & 6  &38.410& 37.522 & 0.68& R & INT \\
           & 26/10/2006 &  7 &38.41&  37.507  & 0.62& R & INT \\
           & 15/12/2006 & 17 &38.423 & 37.639 & 0.90&  & OSN \\
           & 16/12/2006 & 18 &38.423 & 37.648 & 0.91&  & OSN \\
           & 17/12/2006 & 12 &38.424 & 37.658 & 0.93&  & OSN \\
           & 18/12/2006 & 26 &38.424 & 37.668 & 0.95&  & OSN \\
           & 11/01/2007 & 4  &38.430 & 37.974 & 1.31&  & Calar Alto \\
           & 12/01/2007 & 5  &38.430 & 37.989 & 1.32&  & Calar Alto \\
           & 13/01/2007 & 1  &38.430 & 38.004 & 1.33&  & Calar Alto \\
           & 14/01/2007 & 6  &38.430 & 38.019 & 1.34&  & Calar Alto \\
           & 15/01/2007 & 5  &38.431 & 38.034 & 1.35&  & Calar Alto \\
           & 16/01/2007 & 5  &38.431 & 38.050 & 1.36&  & Calar Alto \\
           & 14/09/2007 & 5  &38.491 & 38.020 & 1.33&  & OSN \\
           & 15/09/2007 & 10 &38.491 & 38.008 & 1.32&  & OSN \\
           & 17/09/2007 & 15 &38.492 & 37.980 & 1.30&  & OSN \\
(145486) 2005~UJ$_{438}$
           & 11/01/2007 & 4 &  9.837 & 9.345  & 5.09 &  & Calar Alto \\
           & 12/01/2007 & 4 &  9.834 &  9.359 & 5.14 &  & Calar Alto \\
           & 13/01/2007 & 5 &  9.832 &  9.371 & 5.19 &  & Calar Alto \\
           & 15/01/2007 & 3 &  9.828 &  9.398 & 5.28 &  & Calar Alto \\
           & 16/01/2007 & 7 &  9.826 &  9.410 & 5.32 &  & Calar Alto \\
           & 30/11/2007 & 39 & 9.189 &  8.204 & 0.29 &  & OSN \\
           & 06/01/2008 & 23 &9.123 & 8.372 & 4.16 &  & Calar Alto \\
           & 07/01/2008 & 29 &9.122 & 8.382 & 4.25 &  & Calar Alto \\
           & 26/12/2008 & 15 & 8.594 &  7.635 & 1.54 &   & OSN \\
2007~UL$_{126}$ (or 2002~KY$_{14}$)
           & 01/08/2008 & 15 & 8.665 & 7.793 & 3.62 &  & OSN \\
           & 02/08/2008 & 15 & 8.665 & 7.787 & 3.54 &  & OSN \\
           & 03/08/2008 & 30 & 8.664 & 7.780 & 3.47 &  & OSN \\
           & 04/08/2008 & 25 & 8.664 & 7.774 & 3.40 &  & OSN \\
           & 05/08/2008 & 5  & 8.664 & 7.769 & 3.33 &  & OSN \\
 \hline\hline
\end{longtable}
}

\begin{scriptsize}
\longtab{2}{
\begin{longtable}{llcccccc}
\caption{\label{result} Summary of results from this work. In the table, we present the name of the object, the preferred rotational period
and lightcurve amplitude, the absolute magnitudes (MPC values),
the Julian Date ($\varphi_{0}$) for which the phase is zero in our
lightcurves. Densities are also shown such lower limits (see text).}\\
\hline\hline Object & Designation & Preferred Period [h]&
Amplitude [mag.] & $\varphi_{0}$ [JD]& Absolute magnitude & $\rho$[g/cm$^{3}$] & Binary/Multiple? \\
\hline
\endfirsthead
\caption{continued.}\\
\hline\hline Object & Designation & Preferred Period [h] &
Amplitude [mag.] & $\varphi_{0}$[JD]& Absolute magnitude  & $\rho$[g/cm$^{3}$] & Binary/Multiple? \\
\hline
\endhead
\hline
\endfoot
(55567)~Amycus     & 2002~GB$_{10}$  & 9.76 & 0.16$\pm$0.01 & 2452707.45519 & 7.8 & 0.41 &  \\
(136108)~Haumea    & 2003~EL$_{61}$ &  3.92    & 0.28$\pm$0.02 & 2454112.62040& 0.2  & 2.61 & Yes \\
(136472) Makemake  & 2005~FY$_{9}$   &7.65    & 0.014$\pm$0.002 & 2453796.63861 & -0.3 &  0.66 & \\
(52872)~Okyrhoe    & 1998~SG$_{35}$   & 4.86/6.08    & 0.07$\pm$0.01 & 2454440.62025& 11.3  & 1.65/1.05 & \\
(90482)~Orcus      & 2004~DW    & 10.47    & 0.04$\pm$0.01 & 2453073.36884 & 2.3& 0.35  & Yes \\
(50000)~Quaoar     & 2002~LM$_{60}$  & 17.68/8.84   & 0.15$\pm$0.04 & 2452781.58625
& 2.6 & 0.50 & Yes \\
(42355)~Typhon     & 2002~CR$_{46}$  & 9.67    & 0.07$\pm$0.01 & 2452668.46043 & 7.2 & 0.42 & Yes \\
(20000)~Varuna     & 2000~WR$_{106}$  & 6.3418     &  0.43$\pm$0.01    & 2453376.47462 &  3.6 & 1.03 &  \\
(15874)            & 1996~TL$_{66}$  & 12  &  0.07$\pm$0.02& 2453355.37197 & 5.4 & 0.27 &\\
(12929)            & 1999~TZ$_{1}$  &  10.422   &  0.07$\pm$0.01    & 2454155.67015 & 9.3 & 0.36& \\
(126154)           & 2001~YH$_{140}$  & 13.2 &  0.13$\pm$0.05&  2453355.62794& 5.4 & 0.22 & \\
(55565)            & 2002~AW$_{197}$  &   8.78   &  0.04$\pm$0.01  & 2452672.42954 & 3.3 & 0.50 &   \\
(55636)            & 2002~TX$_{300}$    & 8.16      &  0.04$\pm$0.01 & 2452859.51500 & 3.3 & 0.58 &  \\
(555638)           & 2002~VE$_{95}$  & 9.97 & 0.05$\pm$0.01& 2453024.42248 & 5.3 & 0.39 & \\
(208996)           & 2003~AZ$_{84}$  &  6.79      & 0.07$\pm$0.01  & 2453026.54640 & 3.6 & 0.85 & Yes \\
(120061)           & 2003~CO$_{1}$   &  4.51     & 0.07$\pm$0.01    & 2453024.70117 & 8.9 & 1.92 &  \\
(120132)           & 2003~FY$_{128}$ &  8.54    & 0.15$\pm$0.01  & 2453411.64303 & 5.0 & 0.54 &  \\
(174567)           & 2003~MW$_{12}$  & 5.90/7.87    & 0.06$\pm$0.01   & 2453884.58013 &  3.6 & 1.12/0.63&  \\
(120178)           & 2003~OP$_{32}$  &  4.05      & 0.13$\pm$0.01     & 2453588.39312 & 4.1 & 2.38  &\\
(84922)            & 2003~VS$_{2}$    &   7.42    & 0.21$\pm$0.01     & 2452996.37506 & 4.2 & 0.74& \\
(136204)           & 2003~WL$_{7}$   &  8.24    & 0.05$\pm$0.01     & 2454440.28625 & 8.7 &  0.57& \\
(120347)           & 2004~SB$_{60}$  & 6.09/8.1     & 0.03$\pm$0.01     & 2453588.43205 & 4.4& 1.05/0.59  & Yes \\
(144897)           & 2004~UX$_{10}$  & 5.68    & 0.08$\pm$0.01     &2454358.47542 & 4.7 &1.21  & \\
                   & 2005~CB$_{79}$  & 6.76       & 0.13$\pm$0.02    & 2454472.56600 & 5.0 & 0.86 & \\
(145451)           & 2005~RM$_{43}$  &  6.71     & 0.04$\pm$0.01    & 2454022.46809 & 4.4  & 0.86  & \\
(145452)           & 2005~RN$_{43}$  & 5.62/7.32    & 0.04$\pm$0.01    & 2454358.44257 & 3.9  & 1.23/0.73 &  \\
(145453)           & 2005~RR$_{43}$  &  7.87    & 0.06$\pm$0.01    & 2454031.46931 & 4.0 & 0.63 & \\
(145486)           & 2005~UJ$_{438}$   & 8.32    &  0.13$\pm$0.01     & 2454112.31250& 10.5 & 0.56 &  \\
                   & 2007~UL$_{126}$ or 2002~KY$_{14}$  & 3.56/4.2    & 0.13$\pm$0.01 & 2454680.38646 & 9.4 & 3.09/2.22 & \\
\hline\hline
\end{longtable}}
\end{scriptsize}

\end{document}